# Modeling $p$N$_2$ Through Geological Time: Implications for Planetary Climates and Atmospheric Biosignatures

E.E. Stüeken[1,2,3,4]*, M.A. Kipp[1,4], M.C. Koehler[1,4], E.W. Schwieterman[2,4,5], B. Johnson[6], R. Buick[1,4]

1. Dept. of Earth & Space Sciences and Astrobiology Program, University of Washington, Seattle, WA 98195, USA
2. Dept. of Earth Sciences, University of California, Riverside, CA 92521, USA
3. Department of Earth & Environmental Sciences, University of St Andrews, St Andrews KY16 9AL, Scotland, UK
4. NASA Astrobiology Institute's Virtual Planetary Laboratory, Seattle, WA 981195, USA
5. Dept. of Astronomy and Astrobiology Program, University of Washington, Seattle, WA 98195, USA
6. School of Earth & Ocean Sciences, University of Victoria, Victoria, BC V8P 5C2, Canada
* corresponding author (evast@uw.edu)

**Abstract**

Nitrogen is a major nutrient for all life on Earth and could plausibly play a similar role in extraterrestrial biospheres. The major reservoir of nitrogen at Earth's surface is atmospheric N$_2$, but recent studies have proposed that the size of this reservoir may have fluctuated significantly over the course of Earth's history with particularly low levels in the Neoarchean – presumably as a result of biological activity. We used a biogeochemical box model to test which conditions are necessary to cause large swings in atmospheric N$_2$ pressure. Parameters for our model are constrained by observations of the modern Earth and reconstructions of biomass burial and oxidative weathering in deep time. A 1-D climate model was used to model potential effects on atmospheric climate. In a second set of tests, we perturbed our box model to investigate which parameters have the greatest impact on the evolution of atmospheric $p$N$_2$ and consider possible implications for nitrogen cycling on other planets. Our results suggest that (a) a high rate of biomass burial would have been needed in the Archean to draw down atmospheric $p$N$_2$ to less than half modern levels, (b) the resulting effect on temperature could probably have been compensated by increasing solar luminosity and a mild increase in $p$CO$_2$, and (c) atmospheric oxygenation could have initiated a stepwise $p$N$_2$ rebound through oxidative weathering. In general, life appears to be necessary for significant atmospheric $p$N$_2$ swings on Earth-like planets. Our results further support the idea that an exoplanetary atmosphere rich in both N$_2$ and O$_2$ is a signature of an oxygen-producing biosphere.

**1. Introduction**

Life as we know it is implausible without nitrogen. It is an essential major nutrient for all living things because it is a key component of the nitrogenous bases in the nucleic acids that store, transcribe, and translate genetic information, a necessary ingredient of the amino acids constituting the proteins responsible for most cellular catalysis and at the core of the ATP molecule that is the principal energy transfer agent for biological metabolism. As nitrogen's cosmic abundance is only slightly less than that of carbon and oxygen and because it condenses at moderate distances in circumstellar disks, it should have been available to other potential exoplanetary biospheres. However, N$_2$ gas, the most likely nitrogen species near planetary surfaces in the habitable zone, is nearly inert at standard conditions because of the very strong triple bond in N≡N. Lightning can break this bond and combine nitrogen with other atmospheric species, but on Earth this process has been relatively inefficient (Borucki and Chameides, 1984; Navarro-Gonzalez *et al.*, 2001). Much more significant for removing N$_2$ from Earth's atmosphere is microbial N$_2$ fixation to ammonium, a reaction catalyzed by the enzyme nitrogenase. Today, most of this fixed nitrogen is returned to the atmosphere via the biogeochemical nitrogen cycle, a series of microbially modulated redox reactions that ultimately transform organic nitrogen back to gaseous N$_2$. Thus, life's demand for nitrogen regulates its atmospheric abundance.

Though N$_2$ is not a greenhouse gas itself, its atmospheric partial pressure affects planetary environmental conditions. Higher N$_2$ pressure can enhance greenhouse warming by pressure broadening the absorption bands of such gases as water vapor, $CO_2$, $CH_4$, and $N_2O$ (Goldblatt et al., 2009). Hence, the partial pressure of $N_2$ indirectly influences surface temperature and thus habitability. Moreover, a low total atmospheric pressure of all gases combined weakens the cold-trap for water vapor at the tropopause (Wordsworth and Pierrehumbert, 2014). Where $N_2$ is a major atmospheric constituent, a drop in $pN_2$ can make the tropopause cold-trap leaky (Zahnle and Buick, 2016), allowing water vapor into the upper layers of the atmosphere where it can either remain as vapor, perhaps increasing overall greenhouse warming (Rind, 1998; Solomon et al., 2010; Dessler et al., 2013) (but for a contrary view see Huang et al., 2016), or freeze as high-altitude ice clouds in polar regions, warming the high latitudes and making planetary climate more equable (Sloan and Pollard, 1998). Thus, planetary habitability is dependent, at least in part, on atmospheric nitrogen levels.

Though we know next to nothing about the evolution of the biogeochemical nitrogen cycles on other planets, we now have a better resolved picture of the behavior of nitrogen through Earth's history (Ader et al., 2016; Stüeken et al., 2016; Weiss et al., 2016). It seems that (1) microbial nitrogen fixation evolved very early in Earth's history such that a nitrogen crisis for the primordial biosphere was averted (Stüeken et al., 2015a; Weiss et al., 2016), (2) the partial pressure of atmospheric $N_2$ has fluctuated through time to a greater degree than previously anticipated (Som et al., 2016), (3) an aerobic nitrogen cycle arose before the Great Oxidation Event at ~2.35 Ga (Garvin et al., 2009; Godfrey and Falkowski, 2009), (4) during the mid-Proterozoic aerobic and anaerobic nitrogen cycling was spatially separated under low oxygen conditions (Stüeken, 2013; Koehler et al., in press), and (5) a modern nitrogen cycle with widespread aerobic activity did not arise until the late Neoproterozoic (Ader et al., 2014). The main constraints on these developments were evidently biological evolution and redox changes in Earth's surface environments, principally the oxygenation state of the atmosphere and ocean (Stüeken et al., 2016). Other Earth-like planets may have evolved along somewhat similar pathways with respect to nitrogen cycling, provided that they also originated Earth-like life. If so, then atmospheric swings in $pN_2$ may be a common feature of terrestrial inhabited planets.

In the present study, we investigated the diversity of terrestrial planetary nitrogen cycles by modeling the evolution of Earth's atmospheric $N_2$ reservoir. We then perturbed the model to examine several ahistorical extreme scenarios that could arise on Earth-like exoplanets, defined here as planets with a silicate rock mantle and iron core (empirically < 1.6 Earth radii in size, Rogers, 2015), an orbit within the conservative limits of the habitable zone (Kopparapu et al., 2013), and a similar volatile content to Earth. This conservative definition prescribes a high molecular weight atmosphere dominated by $N_2$, $CO_2$, and $H_2O$ rather than $H_2$ (cf. Pierrehumbert and Gaidos, 2011; Seager, 2013). After exploring a range of variables, we concluded that some combinations of $N_2$ abundance with other gases could act as extraterrestrial biosignatures, others could be "false positives," and yet others may indicate that a planet is uninhabited. Overall, our results suggest that an anaerobic biosphere can greatly facilitate the removal of large amounts of $N_2$ from a planetary atmosphere.

**2. Model setup**
**2.1. Biogeochemical nitrogen box model**

We used the *Isee Stella* software to construct a box model of the global biogeochemical nitrogen cycle (Fig. 1), tracking total nitrogen. Boxes included the atmosphere, pelagic marine sediments deposited on oceanic crust, continental marine sediments deposited on continental shelves and in epeiric seas, continental crust, and the mantle. Continental marine sediments were defined to become part of the continental crust after 100 million years and from there onwards were subjected to metamorphism, which returns nitrogen to the atmosphere. Similarly, we let pelagic marine sediments accumulate for 100 million years before they were subjected to subduction, metamorphism, and volcanism. These timescales were based on the observation that C/N ratios in sedimentary rocks older than about 100 Myr become more variable with higher average values (Algeo et al., 2014). The two sinks of nitrogen from the atmosphere were burial in pelagic sediments and burial in continental sediments with proportions of ~1:25 (Berner, 1982). The sources of atmospheric nitrogen were volcanic and metamorphic degassing of subducted pelagic sediments, metamorphism of continental crust, oxidative weathering of continents, and mantle outgassing. Denitrification was not explicitly included as a source, because we did not track the marine nitrogen reservoir.

The model was run in 1 million-year time-steps from 4.5 Ga to present. Differential equations listed in the appendix were solved with the Euler method. The chosen step size is much lower than the residence time of nitrogen in our modeled reservoirs (>100 Myr), which eradicates computational artifacts that can result in mass imbalances. The initial abundances of nitrogen in the mantle, continental crust, and sediments were set such that the concentrations were the same, assuming that any disequilibrium in concentrations observed today is due to biogeochemical overprinting. The rock masses and modern nitrogen inventories were taken from the work of Johnson & Goldblatt (2015). The initial abundance of atmospheric $N_2$ is unknown. We tuned the model such that the final atmospheric $N_2$ abundance after 4.5 billion years equaled the modern amount of $2.87 \cdot 10^{20}$ mol, defined as one time present atmospheric nitrogen (PAN) (Johnson and Goldblatt, 2015).

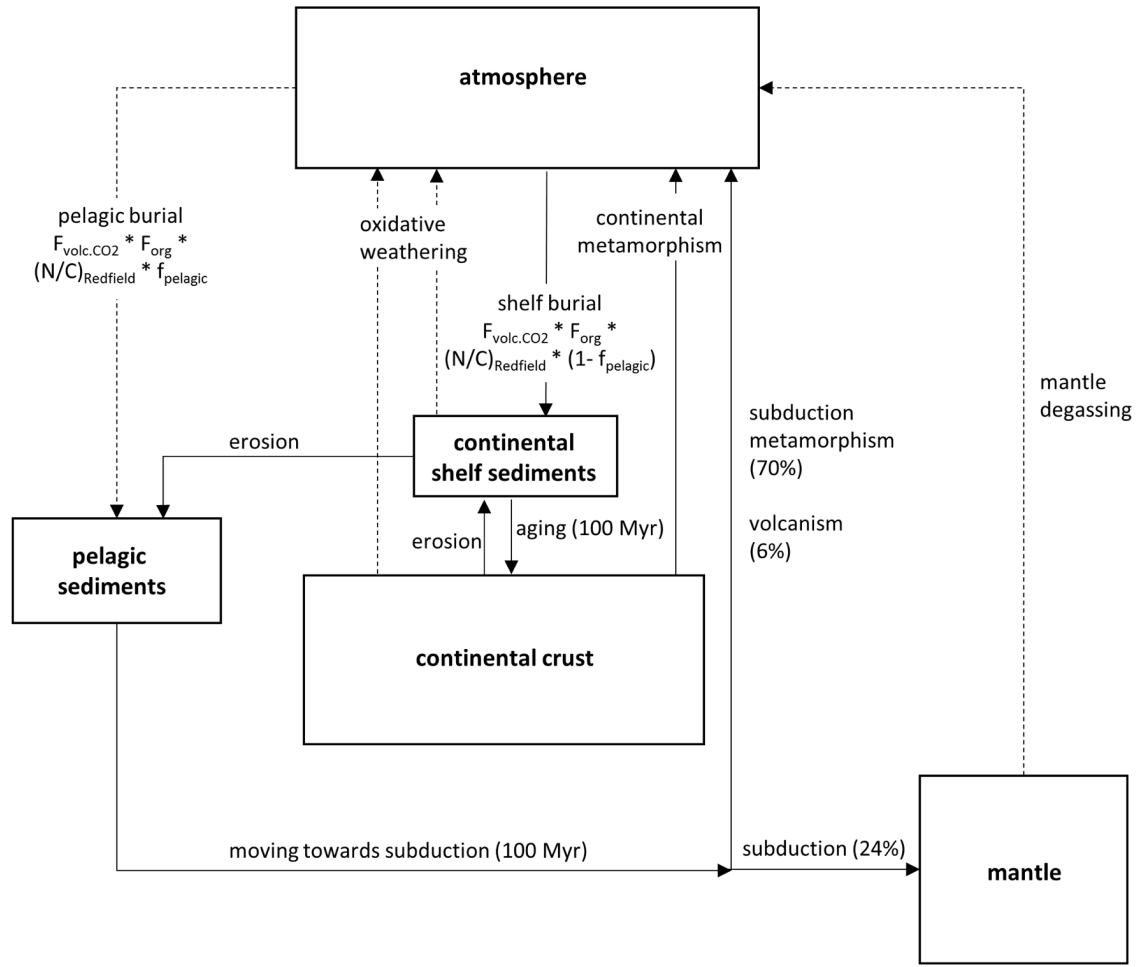

*Figure 1: Schematic of the nitrogen model for the Earth. All fluxes except for burial of nitrogen from the atmosphere into sediments are dependent on the reservoir size. Fluxes with dashed arrows are implemented as time-variable, except in our "base model" (Fig. 2a).*

A more detailed description of how rate constants were derived is given in the appendix. Following the work of Berner (2006b), nitrogen burial was parameterized through biomass burial (*i.e.,* organic carbon). Although this approach neglects the origin and radiation of biological nitrogen metabolisms over Earth's history (Stüeken *et al.*, 2016), it is preferred because (a) it avoids major uncertainties about metabolic rates in deep time, and (b) it is sufficient for tracking the total nitrogen sink from the atmosphere. As further discussed below (Section 4.1), additional nitrogen burial through adsorption on clay minerals is negligible compared to the organic nitrogen flux into sediments. Using this approach, we calculated modern nitrogen burial from the modern volcanic $CO_2$ outgassing flux of $6 \cdot 10^{18}$ mol/Myr (Marty and Tolstikhin, 1998), assuming that 22% ($f_{org}$ = 0.22) of volcanogenic $CO_2$ is buried as organic carbon in the absence of land plants (pre-Devonian) (Krissansen-Totton *et al.*, 2015), and that the C/N ratio of post-diagenetic marine biomass is approximately 10 (Godfrey and Glass, 2011; Algeo *et al.*, 2014). This gave a nitrogen burial flux of $1.32 \cdot 10^{17}$ mol/Myr, equal to ~1.5% of modern biological $N_2$ fixation ($8.64 \cdot 10^{18}$ mol/Myr) (Galloway *et al.*, 2004). We then modulated this flux in deep time in three different ways. First, we took into account secular trends in $f_{org}$ as inferred from the carbon isotope record ("$F_{org}$ model") (Krissansen-Totton *et al.*, 2015). Uncertainties in this record, resulting from potentially underrepresented carbonate reservoirs (Bjerrum and Canfield, 2004; Schrag *et al.*, 2013) and the variance in $\delta^{13}C$ values at any given time point, are discussed below. Second, we assumed a gradual decline in $CO_2$ outgassing from the Hadean to the modern, following the parameterization of Canfield (2004, Equ. 2) ("$F_{org}$ + Heatflow model"). By mass balance, higher $CO_2$ outgassing in the earlier Precambrian implies higher burial fluxes of biomass and with it nitrogen. Uncertainties and caveats of this approach are discussed below. Third, we tested for the effects of additional two-fold increases in $CO_2$ input, and hence nitrogen burial, during superplume events (Abbott and Isley, 2002) ("$F_{org}$ + Heatflow + Superplume model"). This model did not include potential $N_2$ addition through enhanced volcanism, which we will discuss separately. In all of these models, oxidative weathering was implemented as a function of $pO_2$ through time (Lyons et al., 2014) with a reaction order of 0.5, that is, proportional to $(pO_2)^{0.5}$ (Chang and Berner, 1999; Bolton et al., 2006).

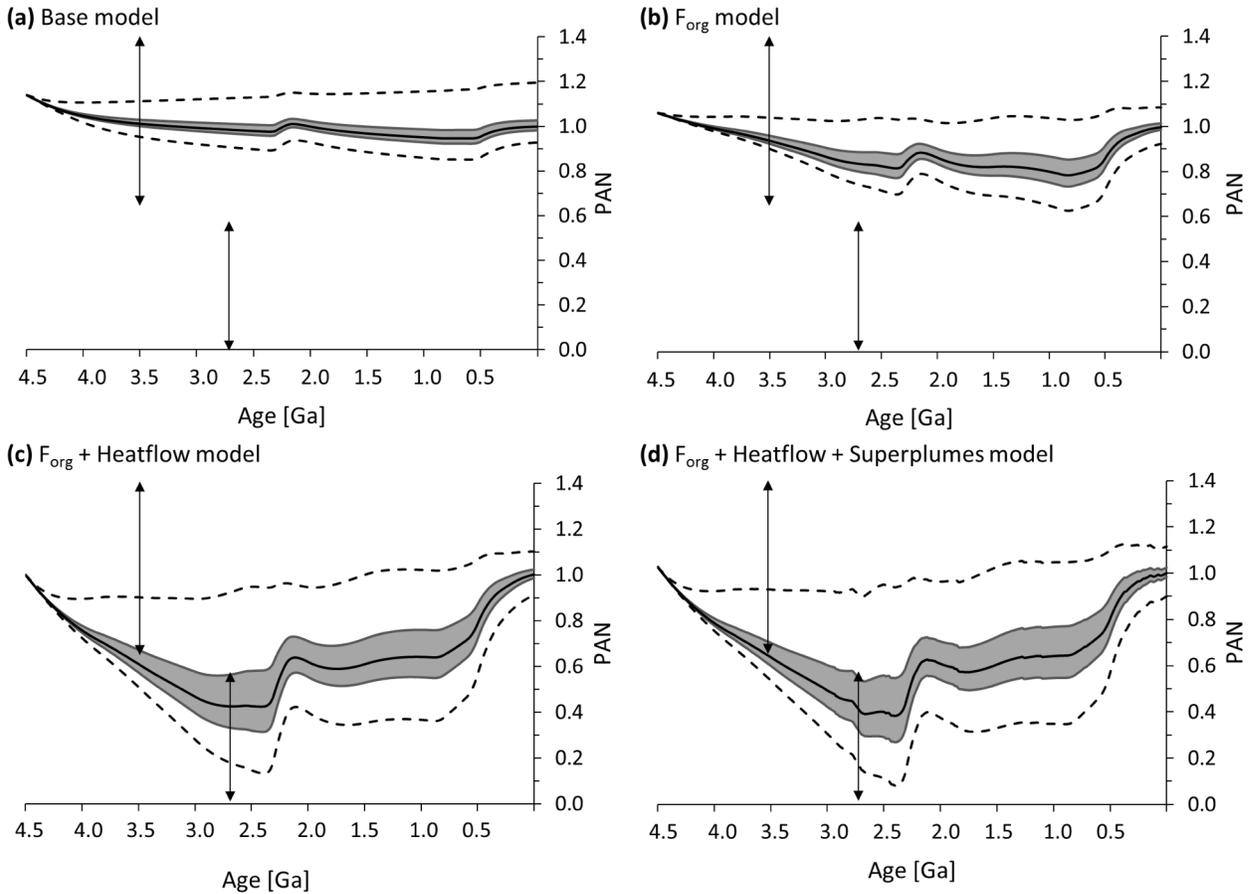

*Figure 2: Reconstructing nitrogen burial over Earth's history.* Arrows mark $pN_2$ value inferred for proxies (Marty et al., 2013; Som et al., 2016). (a) Base model. N burial is held constant, calculated as the product of volcanic $CO_2$ outgassing, the pre-Devonian organic burial fraction ($f_{org}$) of 0.22, and the inverse of the Redfield C/N ratio of 10. (b) $F_{org}$ model. N burial is modulated by secular changes in $f_{org}$ as inferred from the carbon isotope record (Krissansen-Totton et al., 2015). (c) $F_{org}$ + Heatflow model. $CO_2$ outgassing is assumed to have declined gradually from the Hadean to the modern with decreasing heatflow from Earth's interior, as described by Canfield (2004). Carbon burial and with it nitrogen burial is changed in proportion. Mantle outgassing is modulated in the same way. (d) $F_{org}$ + Heatflow + Superplumes model. Additional pulses of $CO_2$ outgassing are assumed during intervals of superplumes as recorded in the rock record (Abbott and Isley, 2002). Solid black line = using best estimates for all parameters; dashed lines = most extreme uncertainty interval if all parameters are off in the same direction (excluding uncertainties in Redfield ratio and modern $CO_2$ outgassing, Appendix A3); grey shaded area = more plausible uncertainty interval with narrower range of values for the metamorphic rate constant (most sensitive variable, see Appendix A3).

In a separate set of models, we tested hypothetical scenarios for Earth-like planets without any biosphere and an anoxic atmosphere, or with a completely anaerobic biosphere that does not experience oxygenation events. Our definition of Earth-like planets includes an orbit within the habitable zone where liquid water can be stable at the surface, as well as a rocky composition and a volatile inventory similar to those of Earth (Section 1). This definition is admittedly limited, but it allows us to identify with greater confidence a subset of parameters that can play a critical role in the history of the nitrogen cycle. We assumed abiotic nitrogen burial was driven by $NH_4^+$ adsorption on clay minerals after abiotic $N_2$ reduction in the atmosphere. Rates were taken from the work of Stüeken (2016) for two extreme end-members corresponding to high and low estimates of abiotic N₂ fixation rates for the Archean Earth and seawater pH of 5 and 8, respectively (see Appendix A1.1). For illustration, these burial fluxes are equal to $1.4 \cdot 10^{-1}$ and $1.7 \cdot 10^{-5}$ times the modern nitrogen burial described above. We did not test abiotic burial under oxic conditions because in that case $NH_4^+$ production would likely be low and nitrogen burial trivial. For the "biotic anoxic" scenario, we used nitrogen burial rates equal to 1-10% of the modern biological N₂ fixation rate. In both the abiotic and biotic anoxic model, oxidative weathering was switched off.

Rate constants for oxidative weathering, metamorphism, erosion, subduction, volcanism, and outgassing were calculated as the ratio of modern fluxes to modern reservoir sizes (see Appendix for details). We further performed simple order-of-magnitude estimates of catastrophic events, such as impacts and large volcanic eruptions, to test whether these could affect our overall conclusions (Section 3.3)

## 2.2. Modeling effects on greenhouse gases & global climate

We selected the most extreme end-members of our Earth model to determine potential effects on Earth's climate. As shown below, the timescales over which variations in $pN_2$ can occur exceed those of the carbonate-silicate feedback cycle (Walker et al., 1981). Hence, changing $pN_2$ alone is unlikely to cause rapid climate changes, because $pCO_2$ can rapidly re-adjust to keep surface temperatures more or less steady. We therefore decided to calculate the required changes in $pCO_2$ that would counterbalance the effects of varying $pN_2$. We used an iterative approach to determine the $pCO_2$ necessary to maintain a 278 K globally averaged surface temperature ($T_{GAT}$), equivalent to the estimated globally averaged temperature during the last glacial maximum. This limit was chosen because geological evidence suggests that the Archean was cool (Hren et al., 2009; Blake et al., 2010) but not permanently glaciated (Young, 1991; de Wit and Furnes, 2016). As input values, we used the changes in nitrogen abundance shown in Section 3 (Fig. 2) and changes in solar luminosity from 3.5 Ga to 2.4 Ga. To do this, we used a 1D radiative convective model that was recently used to calculate habitable zone boundaries (Kasting et al., 1984; Kasting et al., 1993; Kopparapu et al., 2013) and characterize the surface temperature of a hypothetical Archean Earth with a global hydrocarbon haze (Arney et al., 2016). We chose a surface albedo of $A_{surf} = 0.32$, which is a tuning parameter used to reproduce the temperature of modern Earth (Kopparapu et al., 2013; Arney et al., 2016). Our atmospheric composition consisted of only N₂, CO₂, H₂O, and CH₄, where the $pN_2$ was taken from our model output (Table 1) and CH₄ was fixed at $pCH_4 = 0.001$ ($fCH_4$, $fCO_2$, $fN_2$ and the total surface pressure $P_0$ were adjusted self-consistently for changes in $pCO_2$). The H₂O profiles were calculated using a relative humidity profile assuming a surface relative humidity of 80% (Manabe and Wetherald, 1967). The solar luminosity at 3.5 and 2.4 Ga (0.769 and 0.831) was found through interpolation of Table 2 in the work of Bahcall *et al.* (2001). We focused on the $pCO_2$ required to maintain a $T_{GAT}$ > 278 K, noting that significantly lower temperatures ($T_{GAT}$ < 273 K) may allow some open oceans, but a 3D model would be needed to capture their additional

*Table 1: Climate response to changes in atmospheric N₂. Input data are the age, the corresponding relative solar luminosity taken from Bahcall et al. (2001), atmospheric N₂ [PAN] calculated from our box model, and an assumed constant background level of 1000 ppmv CH₄. Partial N₂ pressure (pN₂) was calculated as the product of total N₂ in units of PAN and the modern pN₂ of 0.78 bar. Output parameters are $T_{GAT}$ (global average surface temperature), pCO₂, and P₀ (total average surface pressure, i.e., sum of all gases). Line 1 = starting conditions at 3.5 Ga; line 2 = changing PAN with constant luminosity and constant pCO₂ (taken from line 1); line 3 = changing PAN and luminosity with constant pCO₂; line 4 = changing PAN, luminosity and pCO₂. Parameters are calculated such that the $T_{GAT}$ at (1) and (4) converges to 278 K.*

| line | Age [Ga] | Rel. Solar Luminosity | N₂ [PAN] | $pN_2$ [bar] | $pCO_2$ [bar] | $pCH_4$ [bar] | $P_0$ [bar] | $T_{GAT}$ [K] |
|---|---|---|---|---|---|---|---|---|
| *Forg + Heatflow model:* | | | | | | | | |
| 1 | 3.5 | 0.769 | 0.61 | 0.476 | 0.046 | 0.001 | 0.523 | 278.0 |
| 2 | 2.4 | 0.769 | 0.42 | 0.328 | 0.046 | 0.001 | 0.375 | 274.7 |
| 3 | 2.4 | 0.831 | 0.42 | 0.328 | 0.046 | 0.001 | 0.375 | 282.4 |
| 4 | 2.4 | 0.831 | 0.42 | 0.328 | 0.018 | 0.001 | 0.346 | 278.0 |
| *Forg + Heatflow + Superplumes model, lower limit:* | | | | | | | | |
| 1 | 3.5 | 0.769 | 0.54 | 0.421 | 0.056 | 0.001 | 0.480 | 278.0 |
| 2 | 2.4 | 0.769 | 0.08 | 0.062 | 0.056 | 0.001 | 0.119 | 264.8 |
| 3 | 2.4 | 0.831 | 0.08 | 0.062 | 0.056 | 0.001 | 0.119 | 270.6 |
| 4 | 2.4 | 0.831 | 0.08 | 0.062 | 0.160 | 0.001 | 0.223 | 278.0 |

### 3.1.1. $F_{org}$ model

In the first test, the base model was modified by scaling nitrogen burial as a function of $f_{org}$ (organic carbon burial fraction). This did not significantly change the evolutionary trend of the atmospheric nitrogen reservoir predicted by the base model, as $f_{org}$ has been relatively invariant through geologic time (Fig 2b). During the largest burial event indicated complexities (see Appendix A5 for limitations of our approach). Table 1 shows a summary of our results. We also indicate the surface temperature if $pCO_2$ was not adjusted to maintain $T_{GAT}$ = 278 K, but was maintained at the $pCO_2$ value at 3.5 Ga, and if both the $pCO_2$ and solar luminosity were maintained at the 3.5 Ga values. Table 2 shows similar results as Table 1, but for a more permissive initial and final $T_{GAT}$ of 273 K. This illustrates the sensitivity of the required $pCO_2$ adjustment to compensate for $pN_2$ drawdown as a function of the choice of reference temperature.

**Table 2: Climate response to changes in atmospheric $N_2$.** Same as Table 1, but for a fiducial $T_{GAT}$ of 273.2 K at lines (1) and (4). Differences from Table 1 are in bold.

| line | Age [Ga] | Rel. Solar Luminosity | $N_2$ [PAN] | $pN_2$ [bar] | $pCO_2$ [bar] | $pCH_4$ [bar] | $P_0$ [bar] | $T_{GAT}$ [K] |
|---|---|---|---|---|---|---|---|---|
| *Forg + Heatflow model:* | | | | | | | | |
| 1 | 3.5 | 0.769 | 0.61 | 0.476 | **0.020** | 0.001 | **0.497** | 273.2 |
| 2 | 2.4 | 0.769 | 0.42 | 0.328 | **0.020** | 0.001 | **0.348** | 270.6 |
| 3 | 2.4 | 0.831 | 0.42 | 0.328 | **0.020** | 0.001 | **0.348** | 278.4 |
| 4 | 2.4 | 0.831 | 0.42 | 0.328 | **0.005** | 0.001 | **0.333** | 273.2 |
| *Forg + Heatflow + Superplumes model, lower limit:* | | | | | | | | |
| 1 | 3.5 | 0.769 | 0.54 | 0.421 | **0.024** | 0.001 | **0.446** | 273.2 |
| 2 | 2.4 | 0.769 | 0.08 | 0.062 | **0.024** | 0.001 | **0.087** | 261.1 |
| 3 | 2.4 | 0.831 | 0.08 | 0.062 | **0.024** | 0.001 | **0.087** | 266.2 |
| 4 | 2.4 | 0.831 | 0.08 | 0.062 | **0.118** | 0.001 | **0.181** | 273.2 |

### 3. Results
### 3.1. Nitrogen burial constrained by carbon burial

We tested four different models for the evolution of nitrogen burial through geologic time (Fig 2). In our base model (Fig. 2a), nitrogen burial was held constant at its modern flux. Each subsequent iteration incorporates an additional parameter to the N burial record and, thus, shows additional atmospheric N drawdown. Sensitivity tests are presented in Section A3. The majority of the uncertainty range illustrated in Fig. 2 derives from uncertainties about rates of nitrogen loss during continental metamorphism.

by the carbon isotope record, the Lomagundi Event at ca. 2.3-2.1 Ga, $f_{org}$ temporarily exceeds 0.3 (Precambrian baseline ~0.15-0.2). However, even this increase in organic burial alone appears to be insufficient to significantly deplete the atmospheric $N_2$ reservoir by more than ~0.02 PAN. This model shows a slow depletion of the atmospheric nitrogen reservoir during the Archean (from 1.0 to 0.81 PAN), and relatively constant atmospheric nitrogen during the Proterozoic (range 0.78 to 0.82 PAN). Modern atmospheric nitrogen levels are not attained until late in the Phanerozoic.

### 3.1.2. $F_{org}$ + Heat Flow model

When we scale nitrogen burial proportionally to the amount of $CO_2$ outgassing (Canfield, 2004), our model shows a significant drawdown of atmospheric nitrogen during the Archean, reaching a minimum of 0.44 PAN (= 0.35 bar $N_2$) in the earliest Paleoproterozoic, immediately prior to the Great Oxidation Event (GOE, Fig 2c). Fixed nitrogen is principally buried and stored in continental sediments and continental crust, which reach a maximum of 1.7× the modern continental reservoir size at this time (Section A2). During the GOE, the atmospheric nitrogen reservoir rebounds due to enhanced oxidative weathering of the continents, but this rebound stops after the GOE and atmospheric nitrogen remains low at 0.59-0.65 PAN (= 0.47-0.51 bar $N_2$) until the Neoproterozoic Oxidation Event (NOE). Atmospheric $N_2$ rapidly rises during the Neoproterozoic and Paleozoic in response to a further enhancement of oxidative weathering with the second rise of $O_2$. The later Phanerozoic shows a slow gradual increase in atmospheric $N_2$. As discussed below, we consider this model the most plausible mechanism for explaining a Neoarchean pressure minimum of 0.23 ± 0.23 bar as determined from basaltic amygdales (Som et al., 2016), which is also consistent with basalt amygdale paleobarometry estimates for 2.7 Ga (Som et al., 2016).

To test the plausibility of greater burial enhancement factors, we explored the range from 1× to 10× burial enhancement during plume intervals. A 10× burial enhancement factor for superplume intervals caused the atmospheric nitrogen reservoir to become entirely depleted in the Neoarchean (0.0 PAN at 2.71 Ga), which is implausible

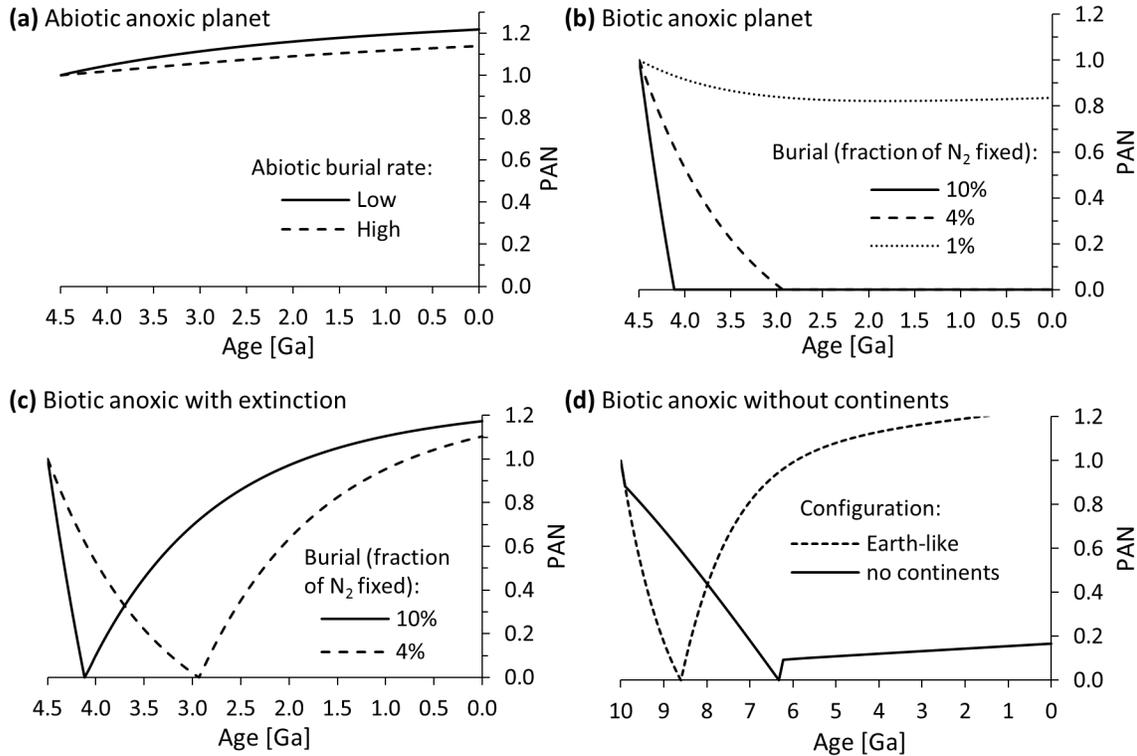

*Figure 3: Hypothetical evolution of atmospheric $N_2$ on extrasolar planets. (a) Abiotic anoxic planet where nitrogen burial is driven by $NH_4^+$ adsorption on clay minerals. High burial rate = $1.5 \cdot 10^{16}$ mol/Myr, low burial rate = $2.2 \cdot 10^{12}$ mol/Myr (Stüeken, 2016). (b) Biological anoxic planet without oxidative remineralization and weathering. Nitrogen burial is calculated as the product of the productivity factor and the modern biological $N_2$ fixation rate on Earth of $8.6 \cdot 10^{18}$ mol/Myr (Galloway et al., 2004). (c) Biological anoxic planet with extinction of the biosphere upon atmospheric loss. (d) Anoxic biological burial and extinction as in panel c with and without burial on continental crust. The burial flux is set equal to 4% of modern biological $N_2$ fixation (dashed line in d describes the same scenario as dashed line in c). Note change in scale on x-axis.*

basaltic amygdales (Som et al., 2016).

### 3.1.3. $F_{org}$ + Heat Flow + Superplumes model

Scaling up N burial by a factor of 2 during plume intervals (Abbott and Isley, 2002) resulted in a small but noticeable effect on the atmospheric nitrogen reservoir (Fig 2d). The effect is most pronounced during the longest superplume event in the geologic record, which is thought to have lasted from 2.775 to 2.696 Ga. Atmospheric nitrogen is drawn down to 0.42 PAN (= 0.33 bar $N_2$) by 2.69 Ga in the wake of this event, which is also consistent with basalt amygdale paleobarometry estimates for 2.7 Ga (Som et al., 2016).

because it would have driven the biosphere to near extinction by removing a major nutrient, or could have induced a global glaciation, for which there is no geological evidence at this time. We thus deemed 10× too extreme of a burial enhancement factor. Values less than 10× and greater than 1× are plausible, but given available data, it is difficult to derive accurate constraints. It is also quite possible that different plume events affected organic burial by different factors. Furthermore, as noted in Section 3.3, nitrogen outgassing from the mantle may have been more pronounced during superplume events, offsetting the effect of enhanced carbon burial. This effect was not considered in our model. All of these complications make assessing the impact of plumes on the atmospheric nitrogen reservoir rather difficult. Still, it is worth noting that, even with a 2× burial enhancement, effects on the atmospheric nitrogen reservoir from plume events are small due to their short duration.

**3.2. Atmospheric nitrogen evolution on abiotic and anoxic worlds**

*3.2.1 Abiotic anoxic planets*

In the hypothetical scenario of a completely abiotic Earth-like planet in the habitable zone of another star with plausible abiotic nitrogen burial rates (Stüeken, 2016), the atmospheric nitrogen reservoir increased slightly from 1.0 PAN to 1.14 -1.22 PAN after 4.5 billion years, with no periods of atmospheric $N_2$ drawdown (Fig. 3a). Catastrophic events such as plumes would have little effect on such a planet because, as further discussed below (Section 3.3.2), most of the nitrogen liberated by such events on Earth is likely sourced from the crust that has been enriched in nitrogen due to biological burial. On an abiotic planet, the crust would be relatively depleted in nitrogen. The sensitivity of these trends to changes in surface and deep Earth nitrogen fluxes is given in the appendix (Section A4). Overall, we find no plausible mechanism that could cause large swings in $pN_2$, apart from the possibility of atmospheric erosion (e.g., Mars, Section 4.3) or freeze-out of $N_2$ on planets far outside of the habitable zone (e.g., Pluto). Although this modeled scenario is hypothetical, it emphasizes the potential importance of life for the evolution of the global nitrogen cycle.

*3.2.2 Biotic anoxic planets*

Our results for a hypothetical Earth-like planet with an anaerobic biosphere suggest that, under the right conditions, biological nitrogen drawdown can have a major effect on the evolution of atmospheric $pN_2$ through time. Nitrogen burial rates equal to 0.1 and 0.04 times the modern $N_2$ fixation rate sequester atmospheric nitrogen to 0 PAN by 4.1 Ga and 2.9 Ga respectively (Fig. 3b). The minimum flux required to reach 0 PAN within 4.5 billion years is roughly 0.03 times modern biological $N_2$ fixation, or 2 times the modern nitrogen burial flux. As in the case of our "$F_{org}$ + Heatflow" model above, most of this nitrogen is stored in continental sediments and crust. A flux of 0.01 times modern biological $N_2$ fixation does not completely draw down atmospheric $N_2$, but leads to a steady state of ~0.84 PAN (Fig. 3b). A more detailed sensitivity analysis of these simulations is presented in the Section A4. When other parameters are set to their most conservative values (*i.e.*, minimizing $N_2$ sequestration), a fixation flux of ~0.13 times modern would be needed to draw down atmospheric $N_2$ to 0 PAN. Again, this scenario is hypothetical, but in comparison to our Earth model, it emphasizes the important influence of atmospheric oxygen on the nitrogen cycle.

It is conceivable that a biosphere would go extinct when the atmospheric $N_2$ reservoir becomes depleted and triggers a global glaciation, at least if biological $N_2$ drawdown does not slow down dramatically as $pN_2$ approaches 0 PAN (e.g. Klingler *et al.*, 1989). Of course, in reality, biogeochemical feedbacks that were not considered in our model may maintain a low, but non-zero, $N_2$ reservoir in the atmosphere, and the biosphere may not necessarily go extinct as evidenced by extreme glacial events on Earth. We nevertheless explored this case in our model, because it provides an estimate of how fast $pN_2$ can recover on an anoxic planet in the absence of an Earth-like atmospheric oxygenation event (*cf.* Figs. 2c, 2d). Our results show that in this case the buried nitrogen would return much more slowly through continental metamorphism and erosion than it does with oxidative weathering. We determined the recovery time of atmospheric $pN_2$ after it is completely sequestered by switching off biological nitrogen fixation once atmospheric $N_2$ reached 0 PAN. With nitrogen burial fluxes of 10% and 4% modern $N_2$ fixation, it takes 2.76 Ga and 2.62 billion years respectively for atmospheric nitrogen values to recover to 1.0 PAN, assuming that the biosphere does not recover during that time (Fig. 3c). This is much slower than the increases in $pN_2$ that occurred in our models over a few hundred million years after the GOE and NOE on Earth, which highlights the linkage between $pO_2$ and $pN_2$ that is further discussed below (Section 4.3).

To assess the effects of continental crust (the major nitrogen repository in our models), we ran a separate model where all burial was directed to pelagic sediments, that is, continents were bypassed to mimic a planet without an equivalent of continental crust. Burial fluxes were arbitrarily set to 4% times the modern fixation rate. Under these conditions, $pN_2$ drawdown to 0 PAN is much slower (Fig. 3d), because nitrogen in pelagic sediments has a much shorter residence time than in continental sediments and crust and is returned relatively rapidly through subduction zone metamorphism and volcanism. However, once atmospheric $N_2$ has been sequestered in the mantle, it does not recover, even over a billion-year timescale. Although this model describes a purely hypothetical scenario of plate tectonics in the absence of continents, it demonstrates that a relatively shallow nitrogen repository akin to continental crust with an intermediate residence time between that of pelagic sediments and the mantle greatly facilitates large atmospheric $pN_2$ swings over hundreds of millions of years, as suggested for the Precambrian Earth (Som *et al.*, 2016).

**3.3. Catastrophic events**

Planets are periodically subject to catastrophic events throughout their history. We performed order-of-magnitude calculations to test whether such events could affect atmospheric N₂ reservoirs and, hence, our overall conclusions. Possible events considered here include asteroid or comet impacts, superplume volcanism as an N₂ source, and large-scale planetary resurfacing.

### 3.3.1. Impacts

There are three ways impacts could affect the amount of nitrogen in a planetary atmosphere: direct nitrogen addition from the bolide, atmospheric erosion with loss of N₂ to space, and release of buried nitrogen through crustal heating. A variety of asteroids have substantial nitrogen contents. The most volatile rich are carbonaceous chondrites, which have 1235 ppm nitrogen on average (Wasson and Kallemeyn, 1988; Johnson and Goldblatt, 2015). If all nitrogen were released as N₂ upon impact, then a hypothetical $3.2 \cdot 10^{21}$ kg of carbonaceous chondrites could contribute one PAN. Such a mass is about 0.05% of Earth's mass, and is approximately equivalent to the once proposed total mass of late heavy bombardment material (Wetherill, 1975). Impacts later in the Archean, post-dating the late heavy bombardment, would have been several orders of magnitude smaller than 0.05% of Earth's mass (Johnson and Melosh, 2012), and hence this nitrogen source was most likely trivial for atmospheric evolution.

Impact erosion has been proposed as an explanation for the thin atmosphere of Mars (Melosh and Vickery, 1989), but more recent calculations suggest that this mechanism may have been less effective than previously thought (Manning *et al.*, 2009). The effect would further decrease with more massive planets than Mars; it is not generally considered to have been significant on early Earth. It is thus probably not a major nitrogen sink on most habitable Earth-like planets.

Large scale crustal heating resulting from impacts could add some nitrogen to the atmosphere (Wordsworth, 2016). Current estimates of nitrogen in continental crust suggest a mass of $1.7 \cdot 10^{18}$ kg, or 0.5 PAN (Goldblatt *et al.*, 2009; Johnson and Goldblatt, 2015). During the late Archean, when atmospheric $pN_2$ may have been as low as 0.44 PAN (Section 3.1.2), the continental reservoir may have been as large as 1.7 times the present continental reservoir, though nitrogen concentrations in continental crust rocks through time are poorly constrained (Section A2). The nitrogen contained in oceanic crust and lithosphere is relatively minor (~0.57 to $0.67 \cdot 10^{18}$ kg N) (Johnson and Goldblatt, 2015). To raise atmospheric $pN_2$ by more than 0.1 PAN through crustal melting, more than 5% of the continental crust would have needed to melt in the Neoarchean and more than 20% in the modern. We consider this unlikely, because there is no geological evidence of such large-scale crustal melting events. We note, however, that some rock types, such as lower continental crust and the oceanic lithospheric upper mantle, are poorly characterized. There are suggestions that parts of the mantle may contain substantial nitrogen (Li *et al.*, 2015), and if a significant fraction of the mantle experienced melting, then large quantities of nitrogen could be added to the atmosphere.

### 3.3.2. Superplumes

Superplume events represent another type of mantle melting that may contribute to changes in atmospheric N₂. While typical mantle melts (*i.e.,* MORB) have a very low nitrogen content of ~1 ppm (Marty, 1995; Marty and Zimmermann, 1999.; Johnson and Goldblatt, 2015), sparse data from continental basalts show higher concentrations. Basalts from the Abitibi region have 6 ppm nitrogen (Honma, 1996), Columbia River basalt has 34 ppm (BCR-1, Govindaraju, 1994), and recent Antarctic basalts around 60 to 100 ppm (Greenfield, 1991). Assuming that most nitrogen will degas during eruption, these concentrations suggest that substantial nitrogen could have been released during flood basalt eruptive events.

Fluid in equilibrium with basaltic melt under oxidizing conditions ($f_{O_2}$ = ΔNNO + 3) has approximately $10^4$ times more nitrogen than the melt itself (Li *et al.*, 2015). Assuming all nitrogen contained in the fluid degasses, an eruption of $3 \cdot 10^{18}$ kg basalt (equivalent to the Siberian Traps) with ~30 ppm nitrogen remaining in the basalt suggests a release of $1 \cdot 10^{18}$ kg nitrogen, or 0.25 PAN. While measurements of nitrogen in flood basalts are quite rare, making this speculative, this simple calculation hints that these events could influence atmospheric N₂ content. However, it is important to note that superplumes in oceanic plates would likely have been much less effective, given that marine basalts tend to have more than one order of magnitude less nitrogen than continental flood basalts (see above); the latter likely assimilate and liberate nitrogen from continental crust. An exception may be oceanic superplumes that pass through parts of the mantle that are enriched in nitrogen due to a subduction overprint. Lamproites and lamprophyres, which are volcanic rocks resulting from the melting of enriched mantle, are notably nitrogen-rich (10s ppm) (Jia *et al.*, 2003), suggesting that such plume events could potentially liberate large amounts of nitrogen into the atmosphere. Nitrogen could also have been liberated by plumes that sampled the lower mantle, which according to some estimates may be nitrogen-rich (Johnson and Goldblatt, 2015). However, both the nitrogen concentration of the lower mantle and the transport pathways of material from such great depth (>660 km) are highly uncertain, making this mechanism difficult to evaluate. Another complication in predicting the effect of superplumes in deep time comes from the possibility that the redox state of magmas may have increased over time, and under more reducing conditions magmatic nitrogen may have been less volatile (Libourel *et al.*, 2003; Kadik *et al.*, 2011; Roskosz *et al.*, 2013; Mikhail and Sverjensky, 2014).

### 3.3.3. Planetary resurfacing

If superplume events are extended to a planetary scale, as is suggested to have happened on Venus, even more nitrogen could be released. The area of the Siberian traps, to continue the above example, is $2.5 \cdot 10^6$ km$^2$, or about 1/200$^{th}$ of the surface of Earth. Multiplying the above estimate for nitrogen released during Siberian Trap volcanism by 200 yields a nitrogen output of $2 \cdot 10^{20}$ kg, or ~50 PAN. Again, we note that this is a highly speculative estimate, but it does suggest the possibility of substantial additions of nitrogen to the atmosphere via large scale volcanism on other planetary bodies that contain substantial amounts of nitrogen in the crust. Overall, catastrophic events could have more marked effects on planets where the crust is nitrogen-enriched, which, as noted above (Section 3.1.2, 3.2.2), is more likely to be the case on planets with a large biosphere that transfers atmospheric nitrogen to crustal repositories.

## 4. Discussion
### 4.1. The evolution of atmospheric $p$N$_2$ on Earth

Although many parameters in the global biogeochemical nitrogen cycle are uncertain and potential reconfigurations of Earth's interior are not taken into consideration for lack of quantitative constraints (Mikhail and Sverjensky, 2014), our results allow us to draw several broad conclusions under the assumption of persistent tectonic cycling through Earth's history as follows:

1. The results from our base model (Fig. 2a), where nitrogen burial is held constant through time while oxidative weathering follows atmospheric $p$O$_2$, show that the oxygenation of the atmosphere alone could probably not have caused the large swings in atmospheric $p$N$_2$ that were suggested by Som *et al.* (2016). Changes in the atmospheric nitrogen reservoir by more than ~0.1 PAN most probably require a change in nitrogen burial over time.

2. Variations in nitrogen burial by up to a factor of 2.9, as inferred from the record of relative organic carbon burial ($f_{org}$), are insufficient to cause significant swings of more than ~0.2 PAN in atmospheric N$_2$ (Fig. 2b). Even if we use $f_{org}$ values from the upper end of the uncertainty range (Krissansen-Totton *et al.*, 2015), the atmospheric N$_2$ reservoir does not drop by more than 0.3 PAN. If $f_{org}$ was smaller than assumed, due to a greater proportion of carbonate formation in oceanic crust or within sediments (Bjerrum and Canfield, 2004; Schrag *et al.*, 2013), this variable would have even less effect on atmospheric N$_2$. To reach atmospheric pressures of less than 0.5 bar at 2.7 Ga (Som *et al.*, 2016), while maintaining pressures of 0.5-1.1 bar N$_2$ at 3.5 Ga (Marty *et al.*, 2013), nitrogen burial must have been markedly higher in the earlier Archean than it is today. There are three possible scenarios to increase nitrogen burial: (i) subduction was more efficient than it is today and metamorphic devolatilization was suppressed; (ii) nitrogen was buried preferentially relative to carbon; (iii) the absolute organic carbon burial flux was much higher, and with it the burial of carbonate, such that $f_{org}$ did not change. It is conceivable that subduction was faster in the Archean (option i), but our sensitivity tests (Section A3) show that shortening the residence time of nitrogen in pelagic sediments by a factor of 2 has minimal effects on the atmospheric reservoir (<0.01 PAN), because the bulk of sedimentary nitrogen is recycled via metamorphism. Metamorphic devolatilization may have been enhanced in the Archean when the geothermal gradient was perhaps somewhat higher than today (Condie and Korenaga, 2010; Cartigny and Marty, 2013), but variations in this parameter also have minimal influence on the output of our model (Section A3). The effects of more rapid subduction and enhanced devolatilization may have more or less canceled each other without a net increase of nitrogen drawdown.

Preferential nitrogen burial (option ii) could potentially occur through adsorption of NH$_4^+$ onto clay minerals. Boatman & Murray (1982) experimentally derived a relationship between the amount of NH$_4^+$ adsorbed on clay and the dissolved NH$_4^+$ concentration in solution. For a doubling of the total nitrogen burial flux, the adsorbed concentration would have to match the concentration of organic nitrogen. Shale samples of sub-greenschist metamorphic grade typically have nitrogen concentrations in the range of 100-1000 ppm with C/N ratios around 40, suggesting that most nitrogen is derived from organics (Stüeken *et al.*, 2016). To achieve this concentration through NH$_4^+$ adsorption alone would require a dissolved NH$_4^+$ concentration of 3-30 mM, which is 30-300 times higher than the NH$_4^+$ concentration of the modern Black Sea (~ 100 µM, Brewer and Murray, 1973), and 100-1000 higher than modern marine NO$_3^-$ levels (~30 µM, Sverdrup *et al.*, 1942). Such high ammonium abundances are also inconsistent with the nitrogen isotope record, which suggests that N-limited ecosystems dominated by biological N$_2$ fixation were initiated in the Mesoarchean at 3.2 Ga (Stüeken *et al.*, 2015a) and persisted until the GOE at ~2.35 Ga (Stüeken *et al.*, 2016). A large reservoir of dissolved NH$_4^+$ should have resulted in isotopic fractionations of up to 27‰ associated with partial NH$_4^+$ assimilation into biomass (Hoch *et al.*, 1992; Pennock *et al.*, 1996), which is not observed. Moreover, due to the extreme energetic cost of splitting the N≡N triple bond, nitrogen fixation should have been suppressed rather than expressed where ammonium was readily available as a nutrient. Enhanced nitrogen burial through adsorption is further inconsistent with the record of C/N ratios, because significant addition of adsorbed NH$_4^+$ would require consistently lower C/N ratios in the earlier Precambrian, which is opposite to observations; C/N ratios tend to increase with increasing age due to preferential nitrogen loss during low-grade metamorphism (Stüeken *et al.*, 2016). NH$_4^+$ adsorption to clays likely did occur during diagenesis, where $NH_4^+$ in pore waters can become enriched to several mM by degradation of organic matter (e.g. Rosenfeld, 1979; Boudreau and Canfield, 1988), but in that case the adsorbed nitrogen is simply returned to the sediment from which it was derived and does not lead to excess nitrogen burial. Enhanced $NH_4^+$ adsorption in the Archean ocean is therefore unlikely to have caused a drawdown in atmospheric $N_2$. Instead, we find it more likely that absolute organic carbon burial, and with it organic nitrogen burial, were significantly higher (option iii).

3. Enhanced volcanic $CO_2$ outgassing in the earlier Precambrian could explain greater nitrogen burial if accompanied by increased burial of both organic matter and carbonate. The observed constancy of $f_{org}$ could have been maintained through carbonatization of oceanic crust as a large carbonate sink (Nakamura and Kato, 2004). Organic matter burial was likely concentrated under anoxic waters along continental margins where sedimentation rates were high. Such an enhanced biomass burial flux would have led to an increase in nitrogen burial and can thus explain the observation of low Neoarchean $pN_2$ (Fig. 2c). Following the formulation of Canfield (2004) for a higher heatflow and proportionally higher volcanic fluxes in deep time, this scenario increases the nitrogen burial flux by 40-80% (depending on the exact age) relative to our early Paleozoic base value, or 3.4-1.9 times above the $f_{org}$ factor alone, throughout most of the Archean. We note that the assumption of a higher Archean $CO_2$ flux (Canfield, 2004; Zahnle et al., 2006) has been challenged by studies arguing for a gradual increase in $CO_2$ outgassing from the Archean into the Proterozoic, concurrent with late continental growth (Holland, 2009; Lee et al., 2016). However, if volcanic $CO_2$ emissions were lower in the Archean than they are today, absolute carbon burial would have been lower and with it the burial of nitrogen. Low Archean $CO_2$ fluxes would only be compatible with high nitrogen burial if $f_{org}$ had been much higher than generally assumed. A high Archean $CO_2$ flux thus remains the most plausible mechanism in our model to explain a decline in $pN_2$ from 0.5-1.1 bar at 3.5 Ga to <0.5 bar at 2.7 Ga, as suggested by paleobarometers (Marty et al., 2013; Som et al., 2016). We will therefore proceed with this assumption, noting that the Archean $CO_2$ flux requires additional constraints to derive a more accurate trajectory for $pN_2$.

We further note that a higher absolute burial flux of organic carbon would constitute a source of oxygen equivalents that would have needed to be balanced by reductants to maintain anoxic surface conditions in the Archean (Kasting, 2013). Proposed fluxes of possible reductants ($H_2$, CO, $H_2S$, $Fe^{2+}$) range over an order of magnitude (reviewed by Zahnle et al., 2006; Holland, 2009) and could therefore plausibly cover the effect of carbon burial. Reductant fluxes may indeed have been higher than previously suggested during the earlier Archean if new evidence for a secular increase in the redox state of Earth's mantle is taken into account (Nicklas et al., 2015). A 2-to-4-fold higher carbon burial flux does therefore not necessarily violate redox balance models.

If we calibrate our model with the results of Som et al. (2016), who inferred an atmospheric pressure of <0.5 bar from the relative sizes of vesicles in basalt flows, then the lower part of our uncertainty range in Fig 2c is more likely to be correct than the upper part. In this case, our model predicts the lowest atmospheric pressure in the Neoarchean and two stepwise increases across the Paleoproterozoic and Neoproterozoic oxidation events, when oxidative weathering progressively shortens the residence time of nitrogen in continental sediments and crust. Our prediction of ~0.6 PAN (= 0.47 bar $N_2$) in the Mesoproterozoic is testable with further analyses of vesicle sizes in Proterozoic lava flows.

4. The effect of superplumes is difficult to assess; they could have led to either more rapid nitrogen recycling through crustal melting or slightly enhanced nitrogen drawdown through carbon burial. The balance may further depend on the redox state of magmas which may have changed over time in favor of progressively more $N_2$ degassing (Mikhail and Sverjensky, 2014). Overall, reconfigurations of the deep Earth are currently poorly constrained, but these could potentially have significant effects, beyond the scope of our model.

### 4.2. Climatic effects of atmospheric $pN_2$ changes in the Archean

Significant changes of >0.1 PAN in our modeled atmospheric $N_2$ abundances occur over several hundred million-year time scales (Fig. 2c,d). Although atmospheric pressure affects the greenhouse efficiency of other atmospheric gases like $CO_2$ through pressure broadening (Goldblatt et al., 2009) and can therefore theoretically cause changes in surface temperature, these time scales are so long that any resulting temperature change could be balanced by the carbonate-silicate feedback cycle, which has a response time on the order of a few hundred thousand years (Sundquist, 1991). As $pN_2$ declines, greenhouse warming decreases, causing the planet to cool. However, with lower temperatures, silicate weathering by carbonic acid slows down, which lowers the sink flux of atmospheric $CO_2$ from the atmosphere (Walker et al., 1981). $CO_2$ would thus build up and balance the temperature change caused by the drop in $pN_2$.

Table 1 shows the required response in $pCO_2$ to our calculated drop in $pN_2$ in the Archean. These calculations also take into account the increasing solar luminosity, which warms the planet and therefore leads to a steady decrease in $pCO_2$. In sum, the effect of rising solar luminosity overpowers the effect of declining $pN_2$ from 3.5 Ga to 2.7 Ga in our nominal model scenario (Fig. 2c), and hence $pCO_2$ would have needed to decrease to maintain a stable surface temperature of 278 K. If $pCO_2$ did not respond, then surface temperature would increase by about 4 degrees over this time interval due to the increase in solar luminosity, despite the drop in $pN_2$. It is only in cases of extreme nitrogen burial, that is, at the lower limit of our uncertainty interval in the model with an additional superplume (Fig. 2d, excluding potential effects of crustal melting), that the decline in $pN_2$ would cause surface temperature to drop by around 7.5 degrees. This drop could have been counterbalanced by a three- to five-fold increase in $pCO_2$. We note that all of our calculated values for $pCO_2$ fall within, or very close to, the range of late Archean $CO_2$ pressures inferred from the rock record (0.003-0.15 bar at 2.5 Ga and 0.004-0.75 bar around 2.7 Ga, Sheldon, 2006; Driese et al., 2011; Kanzaki and Murakami, 2015). Although these estimates vary widely, this agreement suggests that our model results are not unrealistic. Overall, plausible changes in atmospheric $pN_2$ as inferred from our model are unlikely to have resulted in massive climatic perturbations. (We note that a requirement for globally averaged temperatures approaching modern values (~288 K) or higher throughout the Archean would be hard to reconcile with the most restrictive constraints on $pCO_2$ from paleosols even without the climatic impact of falling $pN_2$, which is the well-known Faint Young Sun Paradox. For all but our most extreme scenarios, falling $pN_2$ would only negligibly exacerbate this problem.)

### 4.3. N₂ in extraterrestrial atmospheres

Geological and potential biological processes on other planets may differ markedly from those on Earth, as might the initial volatile inventory. Our results can therefore only provide qualitative trends, but they may nevertheless serve as useful guidelines in evaluating measurements of atmospheric $pN_2$ in exoplanetary atmospheres (Schwieterman et al., 2015b). At the very least, our approach allows us to isolate selected variables that have the potential to play a major role in the evolution of a planet's nitrogen cycle.

Most importantly, nitrogen burial under completely abiotic and anoxic conditions on an Earth-like planet within the habitable zone is likely to be trivial compared to mantle degassing, and hence an uninhabited Earth-like planet with a significant nitrogen inventory is unlikely to ever show low atmospheric $N_2$ pressures (Fig. 3a). This conclusion may be violated in a few cases as follows:

(a) On young, very hot (>1000 K), reducing planets, $N_2$ may be rapidly reduced to $NH_3$ and dissolved in a magma ocean (Wordsworth, 2016). This scenario could probably be ruled out by inferring the planetary temperature through measurements of infrared emission, examination of the planet's atmospheric scale height to determine $H_2$ abundance, and/or observations of the host star to provide an estimate of the planet's age.

(b) Atmospheric $pN_2$ may be permanently low on planets that have lost their atmosphere by erosion and where the outgassing rate is at least an order of magnitude lower so that the atmosphere cannot be replaced (e.g., modern Mars). In this case, however, the abundance of other atmospheric gases would also be very low, and the planet's propensity to lose its atmosphere could be inferred from direct or indirect measurements of its mass and radius and therefore its surface gravity.

(c) Nitrogen burial could be more effective if abiotic $N_2$ fixation by volcanism, lightning, or impacts (Kasting and Walker, 1981; Kasting, 1990; Navarro-González et al., 1998; Mather et al., 2004) is at least an order of magnitude higher than estimated for early Earth. If the pH of the ocean on such a planet is significantly higher than 5, then even larger fixation rates would be required, because otherwise fixed $NH_4^+$ (produced after conversion of $NO_3^-$ to $NH_4^+$ via hydrothermal reduction) (Brandes et al., 1998) would be returned to the atmosphere as $NH_3$ gas instead of adsorbing onto mineral surfaces (Stüeken, 2016). $NH_3$ gas quickly dissociates back to $N_2$ under UV light (Kuhn and Atreya, 1979). So far, a theoretical basis for unusually high extraterrestrial lightning rates is lacking. Enhanced volcanic activity may be detectable remotely through time-dependent observations of sulfate aerosols through transmission spectroscopy (Misra et al., 2015).

(d) Planets may have had a large compositional deficit of nitrogen after the initial period of accretion and enhanced atmospheric erosion by stellar UV light (Lichtenegger et al., 2010; Wordsworth and Pierrehumbert, 2014). This scenario may be detectable through the abundance of other volatiles in the planet's atmosphere or measurements of the nitrogen abundance in the host star (e.g. Brewer et al., 2016).

(e) Planets with a markedly lower oxygen fugacity in their mantle compared to that of Earth may not degas $N_2$, because mantle nitrogen may be stable as $NH_3$ and thus less volatile (Mikhail and Sverjensky, 2014; Li et al., 2015). But such planets may be discernable by the presence of CO rather than $CO_2$ in their atmospheres.

For planets that do not fall within the habitable zone, and thus are not covered by our model results, other scenarios could apply. For example, planets that are closer to the host star than the habitable zone that lack a surface ocean, such as Venus, would show insignificant nitrogen burial, and hence atmospheric $N_2$ would increase as the mantle degasses. This effect would be enhanced on planets with a high oxygen fugacity where $N_2$ outgassing is favored over $NH_3$ retention, as proposed for early Venus (Wordsworth, 2016). Planets far beyond the habitable zone may have low atmospheric $pN_2$ if temperatures drop low enough for $N_2$ to become liquid or solid, such as on modern Pluto and possibly ancient Titan (McKay et al., 1993; Lorenz et al., 1997). Hence, $pN_2$ can fluctuate abiotically in such extreme cases, but for Earth-like planets within the habitable zone as considered in our model, abiotic $N_2$ drawdown is much less likely.

A biosphere on a completely anoxic Earth-like planet can potentially have substantial effects on atmospheric $N_2$ (Fig. 3b). A nitrogen burial flux equivalent to a few percent of modern biological $N_2$ fixation rates (without oxidative remineralization) may be sufficient to deplete the atmosphere of $N_2$ if mantle outgassing rates are comparable to those of Earth. In the absence of oxidative weathering, the only steady return fluxes of buried nitrogen back to the atmosphere would be metamorphism, volcanism, and mantle outgassing, and possibly catastrophic events (Section 3.3.2). It is important to note, however, that burial rates may be significantly different on planets that lack a surface reservoir equivalent to continental crust on Earth (Fig. 3d), which is able to take up and release atmospheric $N_2$ on hundred-million-year timescales.

Overall, our results strengthen the conclusion that the simultaneous presence of significant amounts of *both* $N_2$ and $O_2$ may be a biosignature and indicative of a biosphere with oxygenic photosynthesis (Schwieterman *et al.*, 2015b; Krissansen-Totton *et al.*, 2016) (Fig. 4). As shown above, a large anaerobic biosphere that never "invents" oxygenic photosynthesis can draw down $N_2$ to relatively low levels. Hence, both $O_2$ and $N_2$ would be low. Atmospheric erosion and the possibility of an unusually low mantle fugacity can be evaluated independently in such a scenario (e.g., Mars). On a completely abiotic planet orbiting in the habitable zone a Sun-like star, $O_2$ can build up abiotically, but probably only under the condition that non-condensible gases (including $N_2$) are present in low amounts (Wordsworth and Pierrehumbert, 2014). According to this model, water from a surface ocean would be able to enter the upper atmosphere where it would be photolyzed by UV, causing the H to escape and O to build up after surface sinks for oxidants are depleted. In this case, $N_2$ must start and remain low, otherwise the process is halted. Hence, high levels of abiotic $O_2$ would not co-exist with a thick $N_2$ atmosphere. (Though note the likelihood that abiotic $O_2$ may be substantially higher for planets orbiting M dwarf stars due to other mechanisms not applicable to solar-type host stars (e.g. Harman *et al.*, 2015; Luger and Barnes, 2015). A high-$O_2$-low-$N_2$ scenario is difficult to create biologically given the strong tendency of oxidative weathering and increasing oxygen fugacity in the mantle to release $N_2$ to the atmosphere. An inhabited planet whose biosphere invents oxygenic photosynthesis could eventually transition to oxidative weathering, thereby initiating rapid recycling of buried nitrogen from continental crust as on early Earth (e.g. Fig. 2c). This is thus perhaps the only case where both $N_2$ and $O_2$ reach high relative abundances in the atmosphere. In summary, (1) a planet with high $pN_2$ and no $O_2$ probably has either no biosphere (*e.g.*, Venus) or a very small and/or young biosphere (e.g., first life on the Hadean Earth) that is incapable of transferring large quantities of nitrogen to the crust, (2) a planet with $O_2$ but no $N_2$ may be uninhabited, (3) a planet with neither $O_2$ nor high (modern) abundances of $N_2$ may host an anaerobic biosphere as exemplified by Archean Earth, provided that atmospheric erosion can be ruled out (cf. Mars), and (4) a planet with both significant $N_2$ and $O_2$ suggests the presence a biosphere powered at least in part by oxygenic photosynthesis as on modern Earth. Low total $N_2$ on an anoxic planet (case 3) may be a weak biosignature, which could be confirmed through the detection of other biosignature gases or surface features (e.g. Des Marais *et al.*, 2002; Pilcher, 2003; Domagal-Goldman *et al.*, 2011; Schwieterman *et al.*, 2015a).

## 5. Conclusion

The wide range of uncertainties in all our models, in particular about anything related to possible reconfigurations of Earth's mantle (Mikhail and Sverjensky, 2014), prohibits firm conclusions. Nevertheless, our results allow us to isolate a few key parameters for the evolution of a planet's nitrogen cycle and to formulate hypotheses about the evolution of atmospheric $N_2$ reservoirs on Earth and other planets. Some of these hypotheses may be testable with more constraints on nitrogen fluxes and with additional measurements of geological proxies for atmospheric pressure (Som *et al.*, 2012; Marty *et al.*, 2013; Glotzbach and Brandes, 2014; Kite *et al.*, 2014; Som *et al.*, 2016) (see Kavanagh and Goldblatt, 2015 for possible complications).

To first order, our results suggest that the greatest variability in atmospheric $pN_2$ over the history of a planet can be achieved if the planet is inhabited, if biomass burial is highly variable, and/or if it experiences oxygenation events or large-scale crustal melting. Other abiotic scenarios could be envisioned that could potentially lead to $p$N$_2$ fluctuations, such as N$_2$ freeze-out or atmospheric loss, but many of these cases would likely be discernable through other observations, in particular the orbit of the planet and the abundances of other gases. On an inhabited planet, variation in biomass burial can result from changes in the supply of metabolic substrates including CO$_2$ (as on Earth, Fig. 2c,d) and N$_2$ (as in potential exoplanets, Fig. 3b). In the case of Earth, enhanced biomass burial in the Archean, followed by a stepwise shortening of the crustal residence time across the Paleoproterozoic and Neoproterozoic increases in oxidative weathering, could explain the drawdown and recovery of atmospheric N$_2$ inferred from abundances of N$_2$ in fluid inclusions at 3.5 Ga (Marty *et al.*, 2013) and the size distribution of basaltic amygdales at 2.7 Ga (Som *et al.*, 2016). We note that there is no independent evidence of enhanced burial of both organic carbon and carbonate in the Archean, because total organic carbon (TOC) and carbonate contents of Archean sedimentary rocks are not known to be particularly high. This discrepancy may suggest that large amounts of carbon-rich sediments and carbonated basalts have been subducted and lost. If one were to reject our hypothesis of enhanced nitrogen drawdown into continental crust as a temporary reservoir, then another alternative possibility for explaining a low Neoarchean N$_2$ pressure (Som *et al.*, 2016) would be a much lower initial $p$N$_2$ value in the early Archean, followed by a marked increase in mantle degassing at some time between the Neoarchean and the modern. A test with our "F$_{org}$" model suggests that such a trajectory could be achieved if the mantle degassing rate were an order of magnitude higher throughout Earth's history than thought (Busigny *et al.*, 2011). However, this scenario would be inconsistent with proposed N$_2$ pressures of 0.5-1.1 bar at 3.5 Ga (Marty *et al.*, 2013). There is also no evidence for a major transition in the style or rate of mantle outgassing. But if such a transition occurred, it could conceivably be related to proposed changes in the mantle's redox state (Mikhail and Sverjensky, 2014; Nicklas *et al.*, 2015; Aulbach and Stagno, 2016). Further research is needed to evaluate this possibility.

|  | $pN_2$ | |
|---|---|---|
|  | low | high |
| $pO_2$ low | possibly a (large) anaerobic biosphere (Archean Earth), if atmospheric loss can be ruled out (cf. Mars) | uninhabited (Venus) or a small anaerobic biosphere (possibly Hadean Earth) |
| $pO_2$ high | probably no life | suggests an aerobic biosphere (e.g. modern Earth) |

*Figure 4: Plausible interpretations of observed $N_2$ and $O_2$ abundances in exoplanetary atmospheres. Quantitative constraints on cutoffs for "low" and "high" abundances would need to be evaluated based on measurements of other gases and comparisons to other planets in the observed system.*

Lastly, our results may have some practical implications for observations of extrasolar planets. Despite all the uncertainties in our models, our results suggest that an anaerobic biosphere can – under Earth-like geological conditions – remove significant amounts of N$_2$ from the atmosphere. If multiple terrestrial planets around another star started out with similar volatile contents, but one of them has a significantly lower atmospheric N$_2$ abundance, then this may potentially serve as a biosignature. Measurements of other gases may be necessary to rule out atmospheric erosion as on Mars. In contrast, a planet with an oxygenic biosphere that stimulates oxidative weathering could maintain an atmosphere rich in both N$_2$ and O$_2$, similar to the post-Archean Earth. Our results thus support the idea that the combination of N$_2$ and O$_2$ in an exoplanetary atmosphere may be a signature of a biosphere that is capable of oxygenic photosynthesis (Wordsworth and Pierrehumbert, 2014; Schwieterman *et al.*, 2015b; Krissansen-Totton *et al.*, 2016).

**Acknowledgements**

We thank the NASA postdoctoral program (EES; EWS), the NSF Graduate Research Fellowship program (MAK), the NSF FESD program (grant number 1338810, subcontract to RB), the NSERC Discovery program (BJ), the NASA Exobiology program (grant NNX16AI37G to RB) and the NAI Virtual Planetary Laboratory at the University of Washington (solicitation NNH12ZDA002C and Cooperative Agreement Number NNA13AA93A; EWS, RB) for financial support. We thank Jim Kasting and two anonymous referees for numerous helpful comments that improved the manuscript.

# Appendix to:

# Modeling $pN_2$ through Geological Time: Implications for Planetary Climates and Atmospheric Biosignatures

E.E. Stüeken[1,2,3,4]*, M.A. Kipp[1], M.C. Koehler[1], E.W. Schwieterman[2,4,5], B. Johnson[6], R. Buick[1,4]

1. Dept. of Earth & Space Sciences and Astrobiology Program, University of Washington, Seattle, WA 98195, USA
2. Dept. of Earth Sciences, University of California, Riverside, CA 92521, USA
3. Department of Earth & Environmental Sciences, University of St Andrews, St Andrews KY16 9AL, Scotland, UK
4. NASA Astrobiology Institute's Virtual Planetary Laboratory, Seattle, WA 981195, USA
5. Dept. of Astronomy and Astrobiology Program, University of Washington, Seattle, WA 98195, USA
6. School of Earth & Ocean Sciences, University of Victoria, Victoria, BC V8P 5C2, Canada
* corresponding author (evast@uw.edu)

## A1. Detailed model description

The boxes included the atmosphere, pelagic marine sediments deposited on oceanic crust, continental marine sediments deposited on continental shelves, continental crust, and the mantle. We did not include oceanic crust as a separate box, because nitrogen that is taken up into oceanic basalt during alteration is thought to be derived from sediments (Halama *et al.*, 2014) and was therefore already accounted for by pelagic deposition. Nitrogen burial removes nitrogen from the atmosphere while metamorphism, volcanism, and mantle outgassing return nitrogen. A sensitivity analyses of our model is presented in Sections A3 and A4.

### A1.1. Nitrogen burial

In most of our models, the N burial flux was calculated based on organic C burial, consistent with the approach of Berner (2006b). The total volcanic $CO_2$ flux today ($F_{CO2}(modern)$) is ~$6·10^{18}$ mol/Myr (range $4·10^{18}$ mol/Myr to $10·10^{18}$ mol/Myr) (Marty and Tolstikhin, 1998). In the early Paleozoic, prior to the evolution of land plants, ~22% of that flux was buried as organic carbon ($f_{org}(Paleozoic)$) (Krissansen-Totton *et al.*, 2015). (Land plants are excluded to make our model applicable to all of Earth's history). Multiplying these numbers yields an organic carbon burial flux of around $1.3·10^{18}$ mol/Myr. Average algal biomass has a C/N ratio of around 7 ($(C/N)_{algal}$) (Godfrey and Glass, 2011), while marine sediments from the last few million years display an average C/N ratio of ~13 (Algeo *et al.*, 2014). We assumed a mean value of 10 and calculated a N burial flux of $1.3·10^{17}$ mol/Myr. Holding this value constant through time constitutes our 'base model' (Fig. 2a). The modern nitrogen burial flux ($F_{buial}$) is thus calculated as:

$$F_{burial}(modern) = F_{CO2}(modern) * f_{org}(Paleozoic) * (C/N)_{algal} \qquad (Equ. A1)$$

We then used the record of organic carbon burial ($f_{org}(t)$) to calculate nitrogen burial in deep time ($F_{burial}(t)$). In a first test, we held $CO_2$ input ($F_{CO2}$) constant at the modern value and used the carbon isotope record (Fig. A1) (Krissansen-Totton *et al.*, 2015) to modulate the pre-Devonian organic burial fraction of ~0.22, *i.e.* the variable $f_{org}$ in Equ. (A1) became a function of time (($f_{org}(t)$). This was termed the '$F_{org}$' model (Fig. 2b).

In a second test ('$F_{org}$ + Heatflow'), we assumed higher $CO_2$ input in the Precambrian, as suggested by a hotter mantle, faster plate tectonics and more vigorous volcanism (Sleep and Zahnle, 2001). This made $F_{CO2}$ in Equ. (A1) a function of time ($F_{CO2}(t)$). Greater $CO_2$ input would have raised the burial fluxes of both organic carbon and carbonate without leaving a signature in the $\delta^{13}C$ record. We used the parameterization of Canfield (2004, Equ. 2), which predicts ~3 times higher $CO_2$ input at 4 Gyr and ~2 times higher input at 2.7 Gyr.

In a third test ('$F_{org}$ + Heatflow + Superplumes'), we increased $CO_2$ input and hence nitrogen burial even further during intervals of plume volcanism. We increase N burial by a factor of 2 during each discrete plume event, whose timing was determined from the compilation of Abbot and Isley (2002). In mathematical terms, the time dependence of $F_{org}(t)$ was thus changed slightly relative to the '$F_{org}$ + Heatflow' model. The factor of 2 burial enhancement was suggested for the mid-Cretaceous eruptions of oceanic plateaus and related continental breakup (Caldeira and Rampino, 1991), and thus may not be applicable to superplume events. We therefore also tested larger burial enhancement values to determine maximum plausible effects on the nitrogen cycle (see Section 3.1.3).

In a separate set of models, we explored the effects of nitrogen burial on a completely abiotic planet and on an anoxic planet with a small biosphere. In both cases, oxidative weathering was turned off to mimic the absence of biogenic $O_2$. Nitrogen burial fluxes on an abiotic planet were taken from Stüeken (2016) and kept constant through time. Nitrogen can be fixed abiotically by lightning, impact shock heating and volcanism, which generate $NO_3^-$ (Kasting and Walker, 1981; Kasting, 1990; Navarro-González *et al.*, 1998; Mather *et al.*, 2004). Hydrothermal circulation can convert $NO_3^-$ to $NH_4^+$ (Brandes *et al.*, 1998), which can in turn be buried in sediments by adsorption to clay minerals (Boatman and Murray, 1982). Abiotic nitrogen burial in sediments depends on the pH of seawater, because dissolved $NH_4^+$ dissociates to volatile $NH_3 + H^+$ with a pKa of 9.25 at standard conditions (Li *et al.*, 2012), which lowers the flux of $NH_4^+$ into sediments with increasing pH. We chose two endmembers, one at pH 5 with an $N_2$ fixation flux of $10^{11}$ mol/Myr, and one at pH 8 with an $N_2$ fixation flux of $10^{10}$ mol/Myr. These endmembers correspond to $NH_4^+$ burial fluxes of $1.5 \cdot 10^{16}$ mol/Myr and $2.2 \cdot 10^{12}$ mol/Myr, respectively (Stüeken, 2016).

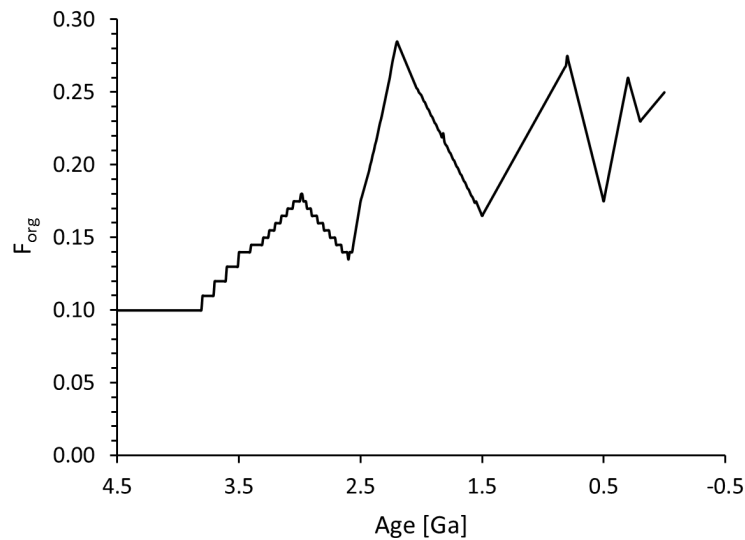

*Figure A1: Reconstruction of $F_{org}$ used in our model (Krissansen-Totton et al., 2015). $F_{org}$ is defined as the fraction of $CO_2$ that is buried as organic carbon rather than carbonate.*

For an anoxic biological planet, we calculated nitrogen burial from the modern biological $N_2$ fixation rate of $8.68.6 \cdot 10^{18}$ mol/Myr (Galloway *et al.*, 2004), multiplied by a productivity factor that was varied from 1% to 10%. We assumed that recycling of fixed nitrogen back to the atmosphere would be negligible in the absence of oxygen, because sulfate and iron oxides are not strong enough oxidizers of ammonium under most environmental conditions (Stüeken *et al.*, 2016). When atmospheric $pN_2$ dropped to zero, we ran separate tests where the biosphere was turned off at that point, mimicking an extinction. For all abiotic and anoxic biological model runs the initial $N_2$ inventory of the atmosphere was set equal to 1.0 PAN. All other parameters (metamorphism, subduction, etc.) were kept at Earth conditions. This is of course a simplification, because these parameters could be markedly different on other planets, but testing a wider parameter space is beyond the scope of this paper. Our chose approach allows us to isolate the effect of life and oxygen.

### A1.2. Shelf versus pelagic burial

In all our biological models, we assumed that 96.2% of biomass burial occurred on continental shelves, while 3.8% occurred in pelagic sediments in the open ocean, *i.e.* in a ratio of 25.3 (Berner, 1982). Burial in the two reservoirs are thus defined as:

$$F_{burialContinent} = F_{burial} * 0.962 \quad \text{(Equ. A2)}$$
$$F_{burialPelagic} = F_{burial} * 0.038 \quad \text{(Equ. A3)}$$

where $F_{burial}$ is defined as in Equ. (A1) or kept constant as in the abiotic and anoxic models.

Today, continental shelves are major locales of carbon burial because productivity is stimulated by nutrients from land and sedimentation rates are high (Berner, 1982). Both factors would also have been valid in the Precambrian. Furthermore, continental shelves in the Archean would have been relatively more anoxic than today, so a larger proportion of fixed carbon and nutrients may have been retained instead of being transported into the open ocean. As shown by Lyons *et al.* (2014), Archean shales in general, including those from continental margins (e.g. Eigenbrode and Freeman, 2006), commonly have TOC levels of several percent, equivalent to those of Neogene anoxic basins and high-productivity areas. These concentrations are significantly higher than the average of ~0.6% of modern continental shelves (Berner, 1982). It is therefore conceivable that organic carbon burial on continental shelves in the Archean was proportionally higher than it is today, at the expense of pelagic sediments.

On the other hand, the relative proportions of pelagic and shelf burial may have changed with continental growth (Fig. A2), but this effect is likely minor, because shelf area would have grown relatively faster than continental mass. First, the surface/volume ratio is generally larger for smaller objects, and second, with less total continental mass, the likelihood of collisions would have been smaller, meaning that the number of land masses and the relative length of continental margins was likely higher than today. Furthermore, it has been proposed that continental relief increased over time (Rey and Coltice, 2008). If so, then a larger proportion of land masses may have been flooded in the past, making shelf area larger, relative to continental mass. It could at times have been even larger than today. Lastly, the total volume of seawater may have been higher in the Archean (Pope *et al.*, 2012), and mid-ocean ridges may have been longer (Hargraves, 1986), which would further increase sealevel and shelf area. Given the number of uncertainties about the evolution of continental volume, continental relief, ocean volume and nutrient distribution in the ocean, we decided to keep the pelagic burial fraction constant through time.

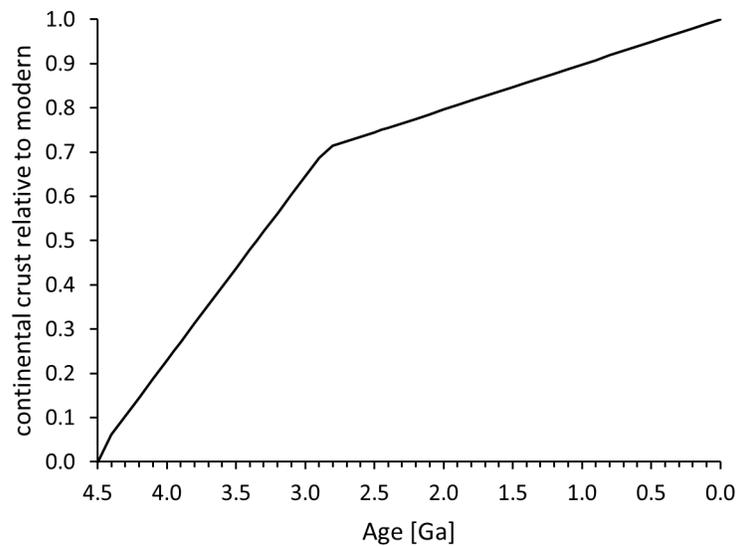

*Figure A2: Evolution of continental volume through time as proposed by Dhuime et al. (2012). Shelf area would have been proportionally higher for smaller continents because of the higher surface/volume ratio of smaller objects and because*

*continents would likely have existed as a larger number of small land masses with potentially lower relief and higher sealevel and ocean volume than today.*

### A1.3. Oxidative weathering

Continental marine sediments and continental crust were subjected to oxidative weathering in our model. We first calculated a rate constant ($R_{wx}$) based on the modern nitrogen weathering flux ($F_{wxCont}(modern)$) relative to the modern continental nitrogen reservoir ($M_{continent}(modern)$). Nitrogen weathering at time $t$ in the past ($F_{wx}(t)$) was then calculated as the product between the rate constant and the nitrogen reservoir in continental crust ($M_{continent}(t)$). The same rate law was applied to continental marine sediments. This flux was further modulated in proportion to atmospheric $pO_2$ with a reaction order of 0.5 (Chang and Berner, 1999; Bolton *et al.*, 2006). Overall, this led to the following expressions for oxidative weathering of continental crust and shelf sediments:

$$F_{wxCont}(t) = R_{wx} * M_{continent}(t) * pO_2(t)^{0.5} \tag{Equ. A4}$$
$$F_{wxSed}(t) = R_{wx} * M_{contSed}(t) * pO_2(t)^{0.5} \tag{Equ. A5}$$

The modern continental reservoir $M_{continent}(modern)$ was assumed to be $1.21 \cdot 10^{20}$ mol (Johnson and Goldblatt, 2015). We used several different sources to determine the modern weathering rate $R_{wx}$. Kump & Arthur (1999) and Berner (2006a) report weathering fluxes for organic carbon. Dividing those by a C/N ratio of 10 and the continental nitrogen reservoir from above yields weathering rate constants for N of 0.00829 $Myr^{-1}$ and 0.0018 $Myr^{-1}$, respectively. Sleep & Zahnle (2001) report a weathering flux for carbonate, but not organic carbon. Assuming a carbonate to organic carbon ratio of about 4:1 in the crust (Berner, 2004) and proportional weathering fluxes, as supported by long-term stability in carbon isotopes (Krissansen-Totton *et al.*, 2015), this yields a nitrogen weathering constant of 0.00351 $Myr^{-1}$, in between the previous two estimates. Alternatively, one could use the modern riverine flux of nitrogen to the ocean (Galloway *et al.*, 2004; Algeo *et al.*, 2014), which yields a weathering rate constant of 0.01478 $Myr^{-1}$ to 0.01657 $Myr^{-1}$. However, the modern flux is likely strongly affected by modern land vegetation and therefore unlikely to be applicable to the Precambrian. We therefore chose 0.00351 $Myr^{-1}$ as a medium value.

As shown by Montross *et al.* (2013), oxidative weathering generates nitrate that is washed into the ocean where it is subject to denitrification together with the marine nitrate reservoir. In our model, which is calibrated by carbon burial, this contribution of fixed nitrogen to the marine biosphere is indirectly accounted for in the net removal flux of nitrogen from the atmosphere through burial. Whatever is not removed, is practically returned to the atmosphere by denitrification. The oxidative weathering flux is therefore directed to the atmosphere in our model.

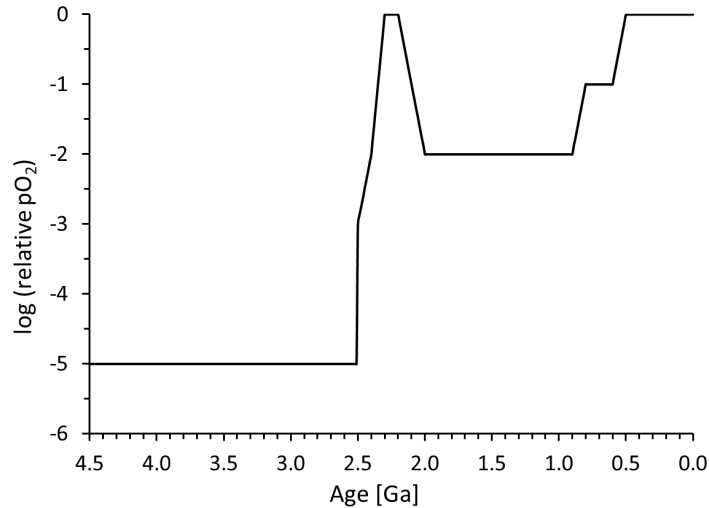

*Figure A3: Evolution of atmospheric $pO_2$ through time used in our Earth models.* The reconstruction of $pO_2$ is taken from Lyons et al. (2014). Oxidative weathering scales with the $pO_2$ with a reaction order of 0.5 (Chang and Berner, 1999; Bolton et al., 2006).

In deep time, we assumed that oxidative weathering scaled with atmospheric $pO_2$ with a reaction order of 0.5 (Chang and Berner, 1999; Bolton et al., 2006). We used the most recent reconstruction of $pO_2(t)$ by Lyons et al. (2014) (Fig. A3). In this reconstruction, $pO_2$ was $10^{-5}$ times present atmospheric levels (PAL) in the Archean (before 2.5 Gyr), it increased to $10^{-3}$ PAL at 2.5 Ga (Kendall et al., 2015), then to modern levels (1 PAL) during the "oxygen overshoot" in the Paleoproterozoic (2.3-2.2 Gyr). It returned to 1% PAL in the Mesoproterozoic (2.0-0.9 Gyr), and rose again to modern levels between 0.8 Gyr and 0.6 Gyr. Transitions between the phases were implemented as linear in logarithmic scale.

## A1.4. Continental erosion and metamorphism

Similar to oxidative weathering, the fluxes for continental erosion ($F_{erosion}$) and in the past at time $t$ were calculated as the product between the continental sediment reservoir ($M_{contSed}$) at time $t$ and a fixed rate constant that was calibrated with the modern. The erosion flux primarily includes turbidites that transport material from the continental shelves to the deep ocean where it can be subducted. To estimate this flux for the modern ($F_{erosion}(modern)$), we calculated the difference between nitrogen added to pelagic sediments by biological productivity today ($5.7 \cdot 10^{15}$ mol/Myr, i.e. 3.8% of all nitrogen that is buried globally, with volcanic $CO_2 = 6 \cdot 10^{18}$ mol/Myr, $f_{org} = 0.25$, and C/N = 10) and the total amount of nitrogen that leaves pelagic sediments via subduction ($5.42 \cdot 10^{16}$ mol/Myr) and metamorphism ($1.63 \cdot 10^{17}$ mol/Myr) (Bebout et al., 1999; Busigny et al., 2003; Busigny et al., 2011). The resulting erosional flux of $2.12 \cdot 10^{11}$ mol/Myr was converted into a rate constant ($R_{erosion}$) of 0.001785 Myr$^{-1}$. This erosional flux implies that pelagic productivity makes up about 0.7% of the total amount of nitrogen in offshore marine sediments, which falls near the lower end of the range of 1-16% derived from a mass balance of organic carbon (Bauer and Druffel, 1998). We therefore considered this the lower limit and tested erosion rates up to 10 times higher. The overall equation for the erosive flux is:

$$F_{erosion}(t) = M_{contSed}(t) * R_{erosion} \tag{Equ. A7}$$

Applying a similar erosive flux from the continental crust into continental sediments, as may occur in rapidly eroding regions that escape oxidative weathering, has minimal effect on the calculated reservoirs sizes (<0.01 PAN).

Nitrogen in continental sediments ultimately gets incorporated into continental crust ("maturation") with a lag time of 100 million years ($T_{Sed}$, tested 50-200 million years). During this lag time, any given parcel of nitrogen that entered this reservoir through burial is subject to oxidative weathering (Section A1.3) and erosion for the duration of the lag time $T_{sed}$. The 'maturation' flux from continental sediments to continental crust was thus defined as:

$$F_{maturation}(t) = F_{burialPelagic}(t-T_{sed}) - [R_{wx} * F_{burialPelagic}(t-T_{sed}) + R_{erosion} * F_{burialPelagic}(t-T_{sed})] * T_{sed} \tag{Equ. A8}$$

Once the nitrogen reached the continental crust, it subjected to metamorphism. By analogy to other fluxes, the continental metamorphic flux ($F_{contMetam}$) at time $t$ was calculated as the product between the crustal reservoir ($M_{continent}$) a rate constant ($R_{contMetam}$) derived from the modern. The rate constant for continental metamorphism ($R_{contMetam}$) was calculated from the flux of organic carbon metamorphism ($4 \cdot 10^{17}$ mol/Myr to $6 \cdot 10^{17}$ mol/Myr) (Wallmann and Aloisi, 2012), an initial C/N ratio of 10 for unmetamorphosed organic matter (Godfrey and Glass, 2011; Algeo et al., 2014) and the assumption that nitrogen escapes 2 times faster than carbon. This approach yielded a rate constant of 0.000829 Myr$^{-1}$ for nitrogen metamorphism. Overall, the continental metamorphic flux was expressed as:

$$F_{contMetam}(t) = M_{continent}(t) * R_{contMetam} \tag{Equ. A9}$$

The assumption of two times faster nitrogen escape relative to carbon is based on the observation that C/N ratios in greenschist to granulite facies rocks display a median of 15 (Stüeken et al., 2015 and references therein). The median increases to 27 if all data points from the Aravalli Group were excluded, which tends to have lower C/N ratios on average compared to other geological units (Papineau et al., 2009; Papineau et al., 2013). Using an average ratio of around 20 compared to an initial ratio of 10 would thus suggest 2 times more nitrogen loss relative to carbon. This number is also supported by the average C/N ratio of 10-30 for total continental crust (1.1-4.0·$10^{21}$ mol C and 1.2·$10^{20}$ mol N) (Hunt, 1972; Berner, 2004; Johnson and Goldblatt, 2015). Therefore, we used a 2x faster nitrogen loss factor in our preferred model. It is true that C/N ratios of low-grade metamorphic rocks can be as high as 100-200 (Stüeken et al., 2015 and references therein), however, at very high grade or during partial melting, relatively more nitrogen is probably retained in silicate minerals while kerogen is preferentially lost. An overall higher loss rate of 2x for nitrogen over carbon therefore seems most appropriate. As upper and lower limits, we consider a maximum uncertainty interval for this number from 1x to 10x, corresponding to scenarios where nitrogen loss occurs at the same rate as carbon loss, or 10 times faster to explain C/N ratios up to 100. This upper limit is likely too high to be a realistic global estimate, but was included for thoroughness in our models. A more likely limit can be derived from the range of median C/N ratios of 15-30, corresponding to 1.5x to 3x faster nitrogen loss relative to carbon. We used this range to arrive at our preferred uncertainty interval.

### A1.5. Subduction zone metamorphism, volcanism & nitrogen sequestration in the mantle

Pelagic sediments were subjected to subduction after their 100 million year transit time ($T_{pel}$) (tested range 50-200 million years). The fate of nitrogen in these sediments was designated as either (1) metamorphic degassing (returned to the atmosphere during metamorphic devolatilization, $F_{pelMetam}$), (2) volcanism (returned to the atmosphere during arc volcanism, $F_{volc}$), or (3) subduction (transferred to the mantle, $F_{sub}$). The relative magnitudes of these fluxes are poorly constrained, so we assembled values from the literature and conducted a sensitivity test to arrive at our preferred values. It is important to note that we applied the same transit time to all three fluxes, *i.e.* we assumed that volcanism, metamorphism and further subduction all operate at the same time, although in reality metamorphism would start several million years earlier. As shown in our sensitivity analysis below, the transit time has minimal effect on the atmospheric $N_2$ budget, and therefore this simplification does not violate our conclusion.

Metamorphic devolatilization in subduction zones constitutes a significant nitrogen loss process (Bebout and Fogel, 1992; Bebout, 1995). In a suite of metasedimentary rocks ranging from low- to high-grade analyzed by Bebout and Fogel (1992), rocks of amphibolite facies were shown to have lost ~80% ($f_{pelMetam}$) of the original nitrogen content as estimated using marine sediments and rocks of lawsonite-albite facies. We therefore explored the range from 70% to 90% for $f_{pelMetam}$, our metamorphic degassing parameter. In mathematical terms, the metamorphic devolatilization flux from subducted marine sediments is thus defined as:

$$F_{pelMetam}(t) = F_{burialPelagic}(t-T_{pel}) + F_{erosion}(t-T_{pel}) * f_{met} \quad \text{(Equ. A10)}$$

The nitrogen remaining after metamorphic degassing either returns to the atmosphere via arc volcanism or it gets subducted into the mantle. The relative proportioning between these two fluxes on a global scale is difficult to assess. Estimates ranging from 100% recycling to the atmosphere (Elkins *et al.*, 2006) to <20% recycling (Mitchell *et al.*, 2010) have been proposed for modern arc settings, with local temperature-pressure regimes appearing to contribute to the large disparities between different subduction zones (Halama *et al.*, 2014). Busigny *et al.* (2011) proposed a 20%/80% partitioning of volcanism/subduction ($f_{volc}/f_{sub}$) as a global budget based on a compilation of several independent studies of subduction zone input and output fluxes. To be thorough, we explored the full range from 0%/100% to 100%/0% volcanism/subduction partitioning in our sensitivity tests. The fluxes for volcanism and subduction are thus defined as:

$$F_{volc}(t) = F_{burialPelagic}(t-T_{pel}) + F_{erosion}(t-T_{pel}) * (1-f_{met}) * f_{volc} \quad \text{(Equ. A11)}$$
$$F_{sub}(t) = F_{burialPelagic}(t-T_{pel}) + F_{erosion}(t-T_{pel}) * (1-f_{met}) * f_{sub} \quad \text{(Equ. A12)}$$

The tests were conducted on our base model without any perturbations to nitrogen burial included. The first criterion that we sought in the model outputs was a secular trend of nitrogen accumulation in the mantle. All models with 90% metamorphic degassing did not meet this criterion. Of the 70% and 80% metamorphic degassing scenarios, all those with less than 30% subduction also did not show secular mantle nitrogen sequestration. From the remaining scenarios, we elected 70% metamorphic degassing, 20%/80% volcanism/subduction as our preferred model because (1) it was the best fit to the global volcanism/subduction estimate from Busigny *et al.* (2011), and (2) it best represented the proposed stability of Phanerozoic $pN_2$.

### A1.6. Mantle outgassing

Estimates for mantle outgassing of nitrogen on the modern Earth ($F_{out}(modern)$) range from 1.4-8.8·$10^{15}$ mol/Myr (Bebout, 1995; Tajika, 1998; Sano *et al.*, 2001; Busigny *et al.*, 2011). We adopted a value of 7.1·$10^{16}$ mol/Myr, since recent studies have favored the higher end of the aforementioned range (e.g. Busigny *et al.*, 2011). We then divided by the size of the modern mantle N reservoir (1.7·$10^{21}$ mol nitrogen, $M_{mantle}(modern)$) (Johnson and Goldblatt, 2015) to calculate an outgassing rate constant ($R_{out}$ = 4.15·$10^{-6}$ Myr$^{-1}$). Mantle outgassing at time $t$ in our model ($F_{out}(t)$) was then calculated by multiplying the rate constant by the size of the mantle reservoir at each time $t$:

$$F_{out}(t) = M_{mantle}(t) * R_{out} \quad \text{(Equ. A13)}$$

We did not vary the rate constant as a function of geologic time in the base model. In models where nitrogen burial was scaled by enhanced $CO_2$ outgassing (following Canfield, 2004, discussed above), the mantle outgassing rate of nitrogen ($R_{out}(t)$) was scaled by the same factor under the assumption that outgassing of $CO_2$ and $N_2$ would follow the same first-order trends. Test runs in which mantle outgassing was not scaled by the same factor as burial showed little deviation from the models that included outgassing modulation, since mantle outgassing is a ~1-2 order of magnitude smaller flux than nitrogen burial.

### A1.7. Differential equations for reservoir evolution

Using the fluxes as defined above, we calculated the rate of change of each nitrogen reservoir ($M_i$) in our model (Equs. A14-A18). Reservoirs are in moles of nitrogen, fluxes in moles per million years.

*Atmosphere:*
$$dM_{atm}/dt = F_{out} + F_{volc} + F_{pelMetam} + F_{contMetam} + F_{wxCont} + F_{wxSed} - F_{burialSed} - F_{burialPelagic} \quad \text{(Equ. A14)}$$

*Continental sediments:*
$$dM_{contSed}/dt = F_{burialContinent} - F_{erosion} - F_{wxSed} - F_{maturation} \quad \text{(Equ. A15)}$$

*Continental crust:*
$$dM_{continent}/dt = F_{maturation} - F_{contMetam} - F_{wxCont} \quad \text{(Equ. A16)}$$

*Pelagic sediments:*
$$dM_{pel}/dt = F_{burialPelagic} + F_{erosion} - F_{pelMetam} - F_{volc} - F_{sub} \quad \text{(Equ. A17)}$$

*Mantle:*
$$dM_{mantle}/dt = F_{sub} - F_{out} \quad \text{(Equ. A18)}$$

The differential equations were solved by the Euler method. Initial reservoir sizes were defined such that the nitrogen concentration was equal in all reservoirs (see main text, Section 2.1). The initial atmospheric nitrogen content was adjusted such that the final results for the modern was equal to 1.0 PAN, except in the abiotic and anoxic models, where the initial value was set to 1.0 PAN.

### A2. Evolution of the continental nitrogen reservoir

Continental marine sediments and continental crust are the major repositories of buried nitrogen in most of our models. Fig. A4 illustrates the nitrogen content of continents as calculated in our "$F_{org}$ + Heatflow" model (Fig. 2b), which is the most plausible model for reconstructing low Neoarchean pressures as inferred by Som *et al.* (2016). The continental reservoir is almost a mirror image of atmospheric $N_2$. Our calculated value for modern crust is within a factor of 2.2 of the amount estimated by Johnson & Goldblatt (2015), based on nitrogen concentrations in different rock types and fractional rock type abundances taken from Wedepohl (1995). Given the range of uncertainties in both our model and the estimates on modern abundances, our results are in fairly good agreement.

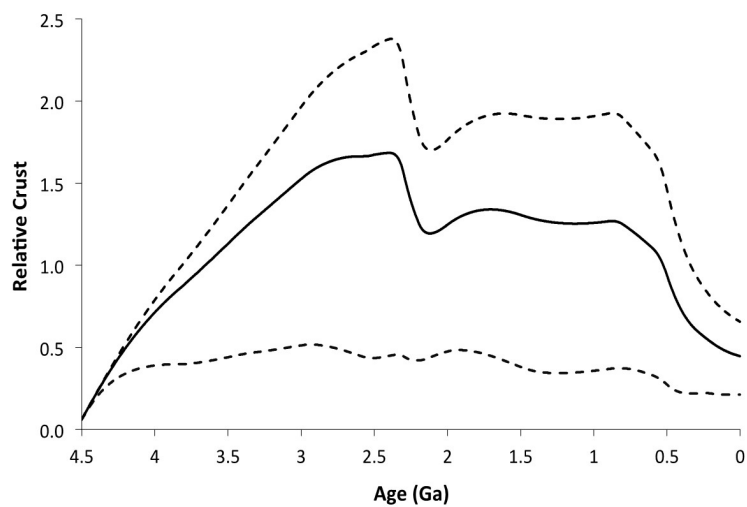

***Figure A4. Nitrogen content of the continental crust through time.*** *Nitrogen accumulates in the continental crust as progressive organic burial sequesters N in this reservoir. The late Archean spike in continental nitrogen corresponds to nitrogen drawdown at this time. Our preferred model predicts a modern continental crust N reservoir size of $5.4\times10^{19}$ mol (i.e. 0.45 times the modern continental reservoir; uncertainty interval $2.6\text{-}7.9\times10^{19}$, i.e. 0.21-0.66 times the modern), which is within a factor of 2.2 of the estimate of $1.2\times10^{20}$ mol by Johnson and Goldblatt (2015).*

## A3. Sensitivity analysis for Earth models

We tested each variable individually to assess its impact on our overall conclusions (Fig. A5 and A6). The curves are compared to our "$F_{org}$ + Heatflow" model, which we consider the most realistic. For most variables, changes of input values within our uncertainty interval have minimal effect (<0.01 PAN) on the results. This includes continental erosion into pelagic sediments (Fig. A5e), the fraction of pelagic metamorphism (Fig. A5h), the fraction of volcanic loss from pelagic sediments (Fig. A6a), and the residence times of sediments on continental shelves (Fig. A6b) and in the deep ocean (Fig. A6c). The initial value of pre-Devonian $F_{org}$ (Fig. A5c) has a minor effect. As shown by Krissansen-Totton et al, (2015), the value is uncertain and possibly variable, but most likely within the range of 0.20-0.22. This range imparts an uncertainty of ~0.06 PAN on the Neoarchean $pN_2$ minimum with no effect on our conclusions.

For the pelagic burial fraction, we tested a range from 1% to 30%. The upper end of this range would only apply in the Hadean and earliest Archean, when continental shelf space may have been lower (Fig. A2), and if modern shelf area is 'maxed out' in terms of carbon burial, such that with smaller continents and constant carbon burial, more burial occurred in pelagic sediments. However, as noted above, high Archean TOC levels in marine shales suggest that Archean shelves were probably capable of holding large amounts of carbon. Furthermore, shelf area grew proportionally faster than continental mass. A smaller uncertainty range for the pelagic burial fraction is therefore more plausible. A range of 1-10% (base value = 3.8%, Berner, 1982) gives an uncertainty of only 0.08 PAN on the Neoarchean $N_2$ minimum (Fig. A5d).

For oxidative weathering, the uncertainty is on the order of 0.2 PAN for weathering rate constants based on organic carbon weathering (see above). A larger rate constant leads to a more dramatic $N_2$ rise with the Paleoproterozoic Great Oxidation Event and a relatively higher atmospheric $N_2$ pressure in the Mesoproterozoic. Even with the lower limit on the rate constant, Neoarchean $pN_2$ would still drop to around 0.54 PAN, equivalent to 0.42 bar of pressure.

Larger uncertainties result from changes in the initial $CO_2$ outgassing rate (Fig. A5a), the C/N ratio of marine biomass (Fig. A5b), and the continental metamorphic rate constant (Fig. A5g). Regarding C/N ratios, there is so far no evidence that they may have changed over the course of Earth's history. We chose a value of 10 as a mean between the canonical Redfield ratio of 7 (Godfrey and Glass, 2011) and observed C/N ratios of 13 in marine mud (compilation by Algeo *et al.*, 2014). In the range of 10-13, the effect on atmospheric $N_2$ is around 0.1 PAN in the Neoarchean. If C/N ratios were significantly different than today, then this may potentially have an impact on our results, as suggested by one model run where the C/N ratio was set to 20 throughout Earth's history (Fig. A5b). In the absence of evidence for changes in C/N ratios, we do not consider this parameter any further in the discussion of temporal changes in $p$N$_2$.

Our modern CO$_2$ outgassing rate of $6 \cdot 10^{18}$ mol/Myr was taken from the most exhaustive and most widely quoted study on this subject (Marty and Tolstikhin, 1998). The authors quote a range from $4\text{-}10 \cdot 10^{18}$ mol/Myr. A higher value leads to a more drastic Archean N$_2$ drawdown in our model (Fig. A5a). Comparing the study by Marty & Tolstikhin (1998) to more recent compilations of CO$_2$ fluxes (e.g. Kerrick, 2001; Berner, 2004, page 59; Fischer, 2008) shows that their estimates are at the middle to upper end. The upper limit of $10 \cdot 10^{18}$ mol/Myr by Marty & Tolstikhin (1998) may therefore be too high. With the more realistic range of $4\text{-}8 \cdot 10^{18}$ mol/Myr the effect on the uncertainty of atmospheric N$_2$ in the Neoarchean is around 0.3 PAN. Using the lower limit of $4 \cdot 10^{18}$ mol/Myr for the modern CO$_2$ flux, Neoarchean $p$N$_2$ would drop to 0.59 PAN, or 0.46 bar, compared to 0.42 PAN (0.33 bar) with a base value of $6 \cdot 10^{18}$ mol/Myr. Because some of our models are already based on higher volcanic CO$_2$ fluxes in the earlier Precambrian, we chose not to consider changes in the modern CO$_2$ flux in the uncertainty interval of Fig. 2 in the main text. The uncertainty about the evolution of CO$_2$ outgassing over 4.5 billion years is likely larger (though unconstrained) than the uncertainty of the modern flux.

The largest impact results from changes in the continental metamorphic rate constant (Fig. A5g). If nitrogen were escaping 10 x faster than carbon, then the Neoarchean $p$N$_2$ minimum would essentially disappear (0.81 PAN, 0.63 bar at 2.7 Ga). As noted above, such a high metamorphic rate is unlikely given C/N ratios of 10-30 for average continental crust and high grade metamorphic rocks. With a more realistic uncertainty range of 1.5-3 times faster nitrogen loss, the uncertainty on atmospheric $p$N$_2$ is roughly 0.2 PAN in the Neoarchean, with a maximum value of 0.56 PAN (0.44 bar) at 2.7 Ga.

In summary, Archean drawdown of atmospheric N$_2$ remains plausible, given the large uncertainty about many of our input variables. Our models do almost certainly not provide accurate reconstructions of $p$N$_2$ through time, but they can convincingly demonstrate that significant swings could have occurred under a reasonable set of conditions.

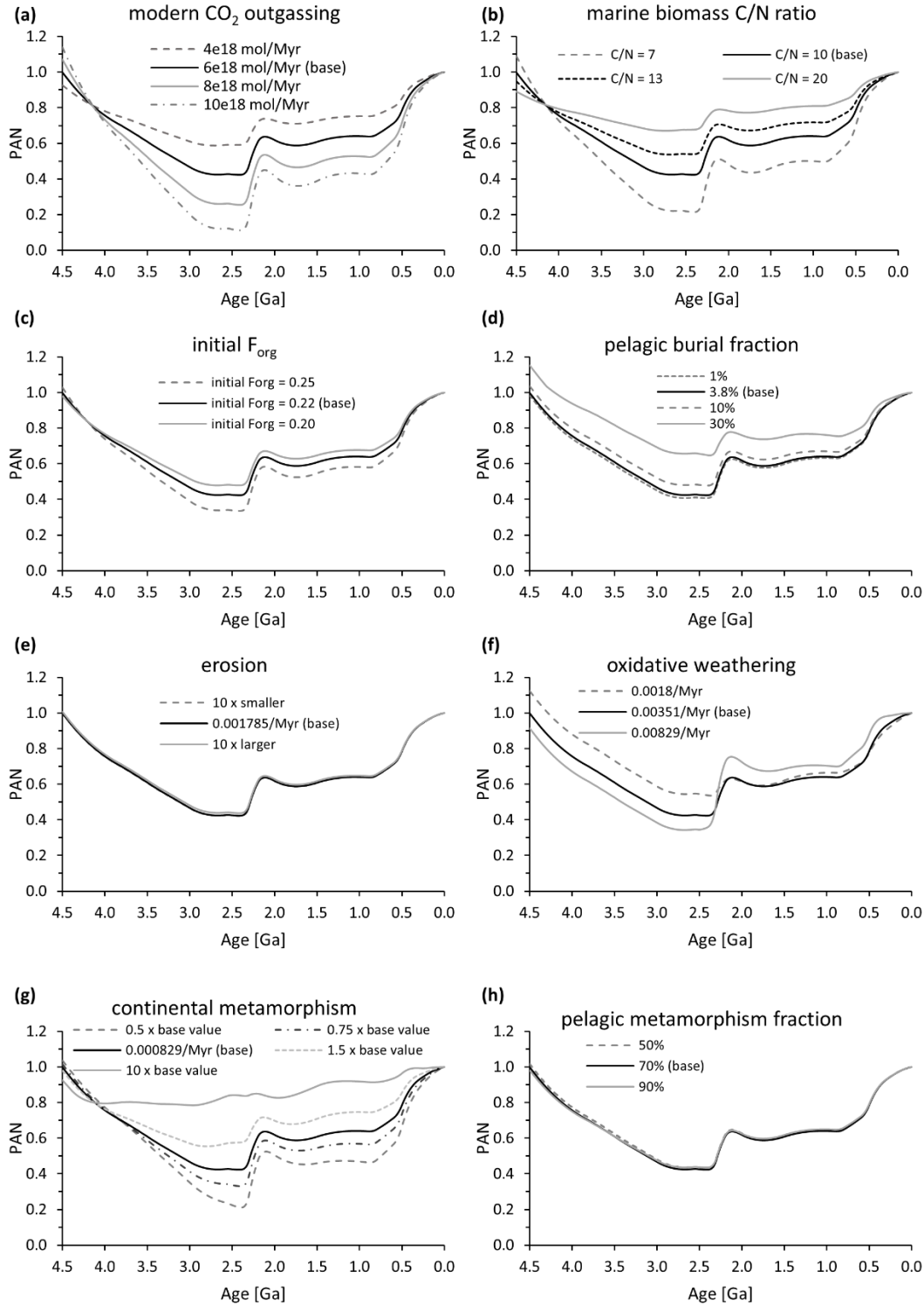

***Figure A5: Sensitivity tests, first batch.*** *Parameters were changed individually while keeping all other parameters equal to base values. Calculations are based on the "$F_{org}$ + Heatflow" model.*

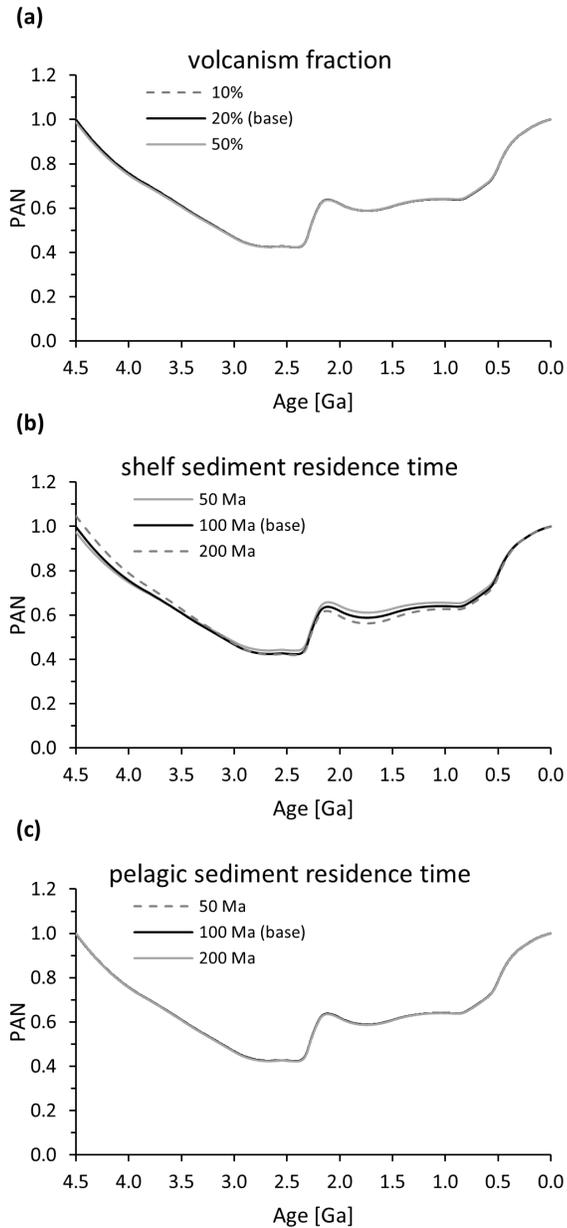

*Figure A6: Sensitivity tests, second batch. Parameters were changed individually while keeping all other parameters equal to base values. Calculations are based on the "$F_{org}$ + Heatflow" model.*

**A4. Sensitivity analysis for hypothetical extraterrestrial models**

We tested the sensitivity of $pN_2$ evolution in our hypothetical extraterrestrial scenarios by varying parameters that were shown to have the most impact in our Earth models. Our extraterrestrial models use estimates of modern biological fixation and abiotic fixation as a base value for the flux of nitrogen out of the atmosphere instead of using $CO_2$ outgassing rates and C/N ratios. Those parameters are therefore not relevant for sensitivity tests. Oxidative weathering is also turned off for these scenarios. So besides varying the total nitrogen fixation rate, we tested the sensitivity to metamorphic rates, erosion, and sediment residence times with the same parameter uncertainties as in Section A3. Similar to our Earth model, atmospheric $pN_2$ turned out to be most sensitive to variations in continental metamorphism. As shown in Fig. A7, the difference in the evolution of the atmospheric N$_2$ reservoir between the preferred scenario and the lower and upper limits increases as the rate of total nitrogen fixation increases. A 0.5 and 5 times change in the continental metamorphism rate does not greatly affect the $p$N$_2$ evolution in our purely abiotic scenarios (~0.002 PAN for the low abiotic burial rate, and ~0.06 PAN for the high burial rate). These differences increase to >0.3 PAN starting at 1% of the estimated modern biologic fixation flux. As noted in the main text, complete N$_2$ drawdown at the upper uncertainty limit would require a burial flux equal to ~13% of modern biological N$_2$ fixation.

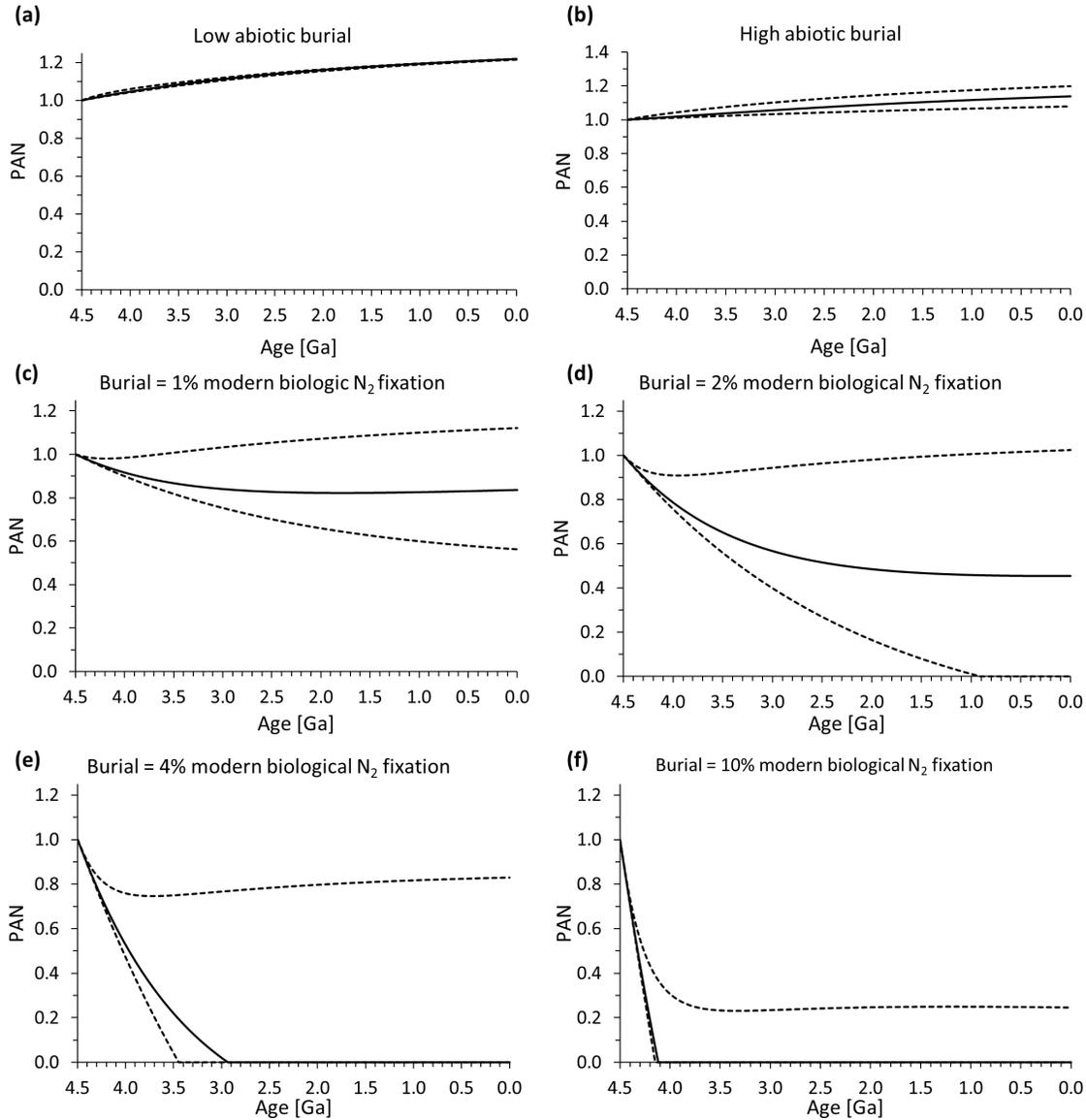

*Figure A7: Sensitivity tests for anoxic and abiotic models. The uncertainty range is primarily controlled by our uncertainty in continental metamorphism.*

**A5. Additional notes on the 1D climate model**

We focused our efforts on conditions that would maintain globally averaged temperatures (T$_{GAT}$) > 278 K (Table 1) and T$_{GAT}$ > 273 K (Table 2). While a T$_{GAT}$ below 273 K has often been assumed to result in a global snowball due to runaway glaciation in 1D radiative climate modeling studies (Domagal-Goldman *et al.*, 2008; Haqq-Misra *et al.*, 2008), recent 3D Global Circulation Modeling (GCM) has shown that a $T_{GAT}$ below 273 K would not necessarily result in global ice cover. For example, Wolf & Toon (2013) found that a substantial (> 50%) open ocean fraction is possible with $T_{GAT} \geq 260$ K. Additionally, Charnay *et al.* (2013) found that a $T_{GAT}$ as low as 248 K can result in an 50° wide ice-free water belt at the equator (see their Figure 9). While 1D climate studies can be sufficient to calculate globally averaged temperatures given a solar luminosity, atmospheric composition, and surface albedo, they cannot by themselves determine whether the planet succumbs to an ice covered state.

Our climatic calculations also may somewhat overestimate the climatic impact of declining $pN_2$. This is because planetary albedo is partially dependent on atmospheric mass through Rayleigh scattering. More massive atmospheres have more Rayleigh scattering and therefore potentially higher albedos, depending on other factors such as surface albedo and cloud properties and fractional coverage. High planetary albedos have a cooling effect on climate. Thus – other factors such as cloud cover being equal – we should expect a warming effect from reduced $pN_2$ that partially counterbalances the cooling effect from reduced pressure broadening of $CO_2$, though this effect is small (Goldblatt and Zahnle, 2011). However, our 1D approach likely underestimates the impact of shortwave scattering on planetary albedo since we parameterize clouds as a high surface albedo. This suggests that our calculations of necessary increases in $pCO_2$ due to declining $pN_2$ are likely upper limits, though 3D GCM modeling may be necessary to confirm this.